\RequirePackage{fix-cm}
\documentclass[smallextended]{svjour3}       
\smartqed  
\usepackage{graphicx}
\usepackage{bm}

\usepackage[squaren]{SIunits}
\usepackage[latin9]{inputenc}

\usepackage{amsmath}
\usepackage{amssymb}

\newcommand{\be}[1]{\begin{equation}\label{#1}}
\newcommand{\ee}{\end{equation}}

\let\pa\partial

\let\r\rho
\newcommand{\rd}{\r_d}

\let\eps\varepsilon

\newcommand{\beq}{\begin{eqnarray}}
\newcommand{\eeq}{\end{eqnarray}}
\newcommand{\beqs}{\begin{eqnarray*}}
\newcommand{\eeqs}{\end{eqnarray*}}
\newcommand{\bequ}{\begin{equation}}
\newcommand{\eequ}{\end{equation}}

\def\r{\rho}

\def\e{\varepsilon}

\newcommand{\bfeta}{{\boldsymbol{\eta}}}

\newcommand{\bfe}{{\boldsymbol{e}}}
\newcommand{\bfF}{{\boldsymbol{F}}}
\newcommand{\bfk}{{\boldsymbol{k}}}
\newcommand{\bfu}{{\boldsymbol{u}}}

\newcommand{\bfv}{{\boldsymbol{v}}}
\newcommand{\bfx}{{\boldsymbol{x}}}

\newcommand{\bfOmega}{{\boldsymbol{\Omega}}}

\newcommand{\D}{{\mathcal D}}

\newcommand{\hsc}{h_{\rm sc}}
\newcommand{\Ro}{\textnormal{Ro}}

\newcommand{\cl}{c_{l}}
\newcommand{\cpd}{c_{pd}}
\newcommand{\Rd}{R_{d}}
\newcommand{\cpv}{c_{pv}}
\newcommand{\Rv}{R_{v}}
\newcommand{\es}{e_{s}}
\newcommand{\eshat}{{\widehat e}_{s}}

\newcommand{\qc}{q_{c}}
\newcommand{\qr}{q_{r}}
\newcommand{\qv}{q_{v}}

\newcommand{\qvs}{q_{vs}}
\newcommand{\qvszero}{q_{vs}^{(0)}}
\newcommand{\qvshat}{{\widehat q}_{vs}}

\newcommand{\phitilde}{{\widetilde\phi}}
\newcommand{\pitilde}{{\widetilde\pi}}

\newcommand{\thetatilde}{{\widetilde\theta}}

\let\pa\partial

\let\eps\varepsilon
\let\eps\varepsilon

\newcommand{\eq}[1]{(\ref{#1})}
\newcommand{\dss}{\displaystyle}
\newcommand{\order}[1]{^{(#1)}}
\newcommand{\ie}{\emph{i.e.}}
\newcommand{\rfr}[1]{#1_{\rm ref}}
\newcommand{\bigoh}[1]{{\mathcal O}\left(#1\right)}
\newcommand{\littleoh}[1]{o\left(#1\right)}

\usepackage{color}
\definecolor{light}{gray}{0.50}
\definecolor{heavy}{gray}{0.35}
\definecolor{black}{gray}{0.0}
\definecolor{dgreen}{rgb}{0.0,0.7,0}
\definecolor{dred}{rgb}{0.9959,0,0}
\definecolor{green}{rgb}{0.0,0.99599,0.0}
\definecolor{purple}{rgb}{0.6,0.0,0.4}

\newcounter{kleincommentno}
\newcounter{hittmeircommentno}
\begin{document}

\title{Asymptotics for moist deep convection~I: \\
Refined scalings and self-sustaining updrafts
}


\author{Sabine Hittmeir        \and
        Rupert Klein 
}


\institute{S. Hittmeir  \at
              University of Vienna, Austria \\
              \email{sabine.hittmeir@univie.ac.at}           
           \and
           R. Klein \at
              Freie Universit\"at Berlin, Germany \\
				\email{rupert.klein@math.fu-berlin.de}}

\date{Received: date / Accepted: date}

\maketitle

\begin{abstract}
Moist processes are among the most important drivers of atmospheric dynamics,
and scale analysis and asymptotics are cornerstones of theoretical meteorology. 
Accounting for moist processes in systematic scale analyses therefore seems of
considerable importance for the field. 
Klein \& Majda (TCFD, {\bf 20}, 525--552, (2006)) proposed a scaling regime 
for the incorporation of moist bulk microphysics closures in multiscale 
asymptotic analyses of tropical deep convection. This regime is refined here 
to allow for mixtures of ideal gases and to establish consistency with a more 
general multiple scales modelling framework for atmospheric flows. 

Deep narrow updrafts, so-called ``hot towers'', constitute principal building 
blocks of larger scale storm systems. They are analysed here in a sample 
application of the new scaling regime. A single quasi-onedimensional
columnar cloud is considered on the vertical advective (or tower life cycle) 
time scale. 
The refined asymptotic scaling regime is essential for this example as it reveals a
new mechanism for the self-sustainance of such updrafts. Even for strongly
positive convectively available potential energy (CAPE), a vertical balance 
of buoyancy forces is found in the presence of precipitation. This balance 
induces a diagnostic equation for the vertical velocity and it is responsible 
for the generation of self-sustained balanced updrafts. The time dependent 
updraft structure is encoded in a Hamilton-Jacobi equation for the 
precipitation mixing ratio. Numerical solutions of this equation 
suggest that the self-sustained updrafts may strongly enhance hot tower life cycles.
\keywords{Moist atmospheric flows
\and multiscale asymptotics
\and matched asymptotic expansions
\and hot towers
\and cumulonimbus clouds}
\end{abstract}


\section{Introduction}
\label{intro}

Solutions to the full governing equations for atmospheric flows are 
analytically and numerically challenging and often difficult to interpret.
Model reductions by scale analysis provide important complementary insights 
and therefore have a long history in meteorology. 
One key technique that allows for systematic studies of complex process 
interactions across disparate length and time scales is multiple scales 
asymptotics. 

Of particular interest in meteorology are multiscale interactions 
involving moist physics processes. A bulk microphysics closure for the 
moisture dynamics was first successfully incorporated into an asymptotics 
framework by \cite{KleinMajda2006}. The warm 
cloud moisture model used in that study consists of balance equations for water 
vapor, cloud water, and rain water, with phase exchange terms corresponding to 
the basic ``Kessler scheme'', \cite{Kessler1969}. 
This type of parameterisation has been widely used in various variants and 
generalizations in meteorological modelling, see also \cite{CottonEtAl2011,%
DurranKlemp1982,Emanuel1994,GrabowskiSmolarkiewicz1996,KlempWilhelmson1978}.

A prerequisite for multiple scales analyses is that all of the 
participating scale-dependent phenomena are represented within one common 
asymptotic scaling regime or distinguished limit \cite{Klein2010}. The moist
process scaling regime introduced in~\cite{KleinMajda2006} is 
suitable for the purpose of studying the short-time evolution of hot convective
cloud towers, but it is not fully consistent with the broadly applicable, unified 
asymptotic modelling framework developed in parallel and summarized in 
\cite{Klein2010}. It also neglects the individual thermodynamic properties of the 
moisture species. 

The first goal of the present work is therefore to redefine 
the asymptotic regime for moist physics so as to reconcile it with the general 
framework and to make it applicable, thereby, to more general atmospheric flow 
situations. At the same time, we allow for more realistic thermodynamics by 
including mixtures of ideal gases and state dependent latent heat. 

Latent heat conversion due to phase changes of water in the atmosphere 
strongly influences its energy balance. Of particular interest here are ``hot 
towers'', \ie, large deep convective cumulonimbus clouds that occupy small 
horizontal scales with typical diameters of a kilometer. It is common belief 
that these hot towers are to a great extent responsible for the vertical 
energy transport to the upper troposphere within the intertropical 
convergence zone (see \cite{RiehlMalkus1958,HoltonHakim2013} and references 
therein). Moreover they are the building blocks of intermediate scale 
convective storms, \cite{LearyHouze1980}.   

Using asymptotic techniques, a multiscale model for the short time evolution 
of hot towers and their interaction with internal waves was developed 
in~\cite{KleinMajda2006}. That work also forms the basis for subsequent 
investigations in~\cite{RuprechtEtAl2010,RuprechtKlein2011}.
These reveal that moisture can reduce the vertical energy transport by 
internal gravity waves, which may be of considerable importance 
for climate models. The internal wave time scale considered in that study
is much shorter, however, than the time scale of vertical convection in a
cloud tower or than the overall cloud tower life cycle time scale. As a 
consequence, the asymptotic regime considered in \cite{KleinMajda2006} is
not appropriate for studying cloud dynamics in the context of, e.g., 
self-organized convection and the formation of strong storm fronts (squall 
lines) or the development of an atmospheric vortex. 

The second goal of the present work is, therefore, to reconsider the asymptotics
of narrow deep convective hot towers on time scales comparable to the vertical
advection time within a tower. An individual, essentially isolated cloud tower 
is analyzed here as an example for applications of the new asymptotic scaling 
regime. Interactions between towers in multicellular convection will be addressed 
in a companion paper. In combination with the refined moist physics closure scheme, 
we find on this larger time scale an interesting and apparently new regime of 
self-sustained precipitation-driven convection.  

The outline of the rest of the paper is as follows. The remainder of this 
introduction briefly summarizes the main results of the paper. 
Section~\ref{sec:GoverningEquations} introduces the governing equations for 
cloudy air, which are then rendered dimensionless in section 
\ref{sec:Nondimensionalisation}. That section also summarizes two distinguished
limits for the moist variables: One is pragmatically defined just on the 
basis of bare magnitudes of various thermodynamics parameters, the other is 
derived from a detailed scale analysis of the hydrostastic moist adiabatic 
distribution in section~\ref{sec:RevisedScalings}. 
The governing equations are reconsidered in section \ref{sec:CloudTowerDynamics}
based on an asymptotic ansatz for narrow cloud towers, and the tower evolution
equations on the convective time scale for saturated and undersaturated regions are 
derived through boundary layer type asymptotic arguments. Slight differences between 
the two asymptotic scaling regimes from section~\ref{sec:Nondimensionalisation} are 
discussed. Section~\ref{sec:UpAndDownDrafts} provides sample numerical solutions of the
new convective time scale tower dynamics equations to reveal the essential 
physical mechanisms they encode. The potential for self-sustainance of precipitating
deep convective updrafts is discussed, in particular. We close with a summary and 
further discussion in section~\ref{sec:Conclusions}.


\subsection{Summary of the main results}


\subsubsection{Asymptotic scaling regimes for moist air thermodynamics}
The thermodynamic characteristics of moist air are captured by the
equation of state parameters summarized in table~\ref{tab:ThermodynamicsParameters}.
Together with a reference temperature, $\rfr{T}$, these are seven dimensional 
quantities of influence involving the two independent physical dimensions of 
specific energy, measured in units of $\joule\per\kilo\gram$, and temperature,
measured in $\kelvin$, and this gives rise to five independent dimensionless
parameters as listed in table~\ref{tab:DistinguishedLimit}. 
\begin{table}
\caption{Thermodynamic equation of state parameters for moist air at reference 
temperature $\rfr{T}=273.15$ K (see, e.g., \cite{CottonEtAl2011,Emanuel1994}):}
\begin{center}
\begin{tabular}{crl@{\quad}l}
$\cpd$
  & $1005$
    & $\joule\per\kilogram\per\kelvin$
      & dry air specific heat capacity at constant pressure 
    \\
$\Rd$
  & $287$
    & $\joule\per\kilogram\per\kelvin$
      & dry air gas constant
        \\
$\cpv$
  & $1850$
    & $\joule\per\kilogram\per\kelvin$
      & water vapor specific heat capacity at constant pressure 
        \\
$\Rv$
    & $462$
      & $\joule\per\kilogram\per\kelvin$
        & water vapor gas constant
          \\
$\cl$
  & $4218$
    & $\joule\per\kilogram\per\kelvin$
      & liquid water specific heat capacity  
        \\
$\rfr{L}$
  & $2.5\cdot 10^{6}$
    & $\joule\per\kilogram$
      & latent heat of condensation at reference conditions
        \\
\end{tabular}
\end{center}
\label{tab:ThermodynamicsParameters}
\end{table}
\begin{table}
\caption{Approximate values and scalings of dimensionless 
thermodynamics parameters. The quantities $\Gamma, k_v, A, k_l, L, E, \kappa_v$
are considered to be constants of order $\bigoh{1}$ as $\eps\to 0$. 
}
\label{tab:DistinguishedLimit}       
\begin{tabular}{lrrr}
\hline\noalign{\smallskip}
Para-
  &  
    & regime
      & regime
        \\
meter 
  & value 
    &  \hbox to 0.9cm{ $\alpha = 0$ }  
      &  \hbox to 0.9cm{ $\alpha = 1$ } \\
\noalign{\smallskip}\hline\noalign{\smallskip}
$\dss \frac{R_d}{c_{pd}}$ 
  & 0.29
    & $\e \Gamma$ 
      & $\e \Gamma$
        \\[10pt]
$\dss \frac{c_{pv}}{c_{pd}}$ 
  & $1.84$
    & $\e^{-1}k_v$
      & $k_v$
        \\[10pt]
$\dss \frac{R_{v}}{c_{pd}}$ 
  & 0.46 
    & $1/A$ 
      & $1/A$
        \\[10pt]
$\dss \frac{c_{l}}{c_{pd}}$ 
  & 4.2 
    &  $\e^{-1} k_l$
      & $\e^{-1} k_l$
        \\[10pt]
\noalign{\smallskip}\hline
\end{tabular}
\hfill
\begin{tabular}{lrrr}
\hline\noalign{\smallskip}
Para-
  &  
    & regime
      & regime
        \\
meter 
  & value 
    &  \hbox to 0.9cm{ $\alpha = 0$} 
      &  \hbox to 0.9cm{ $\alpha = 1$ }  \\
\noalign{\smallskip}\hline\noalign{\smallskip}
$\dss \frac{\rfr{L}}{c_{pd}\rfr{T}}$ 
  & 9.1
    & $\e^{-1} L$ 
      & $\e^{-1} L$
        \\[10pt]
\noalign{\smallskip}\hline\noalign{\smallskip}
derived parameters \span\omit\span\omit \\
\noalign{\smallskip}\hline\noalign{\smallskip}
$\dss \frac{R_{d}}{R_{v}}$ 
  & 0.62
    & $\e E$ 
      & $E$
        \\[9pt]
$\dss \frac{c_{pv}}{c_{pd}}\frac{R_d}{c_{pd}} - \frac{R_v}{c_{pd}}$ 
  & 0.067
    & $\kappa_v$ 
      & $\e \kappa_v$
        \\[9pt]
\noalign{\smallskip}\hline
\end{tabular}
\end{table}

Table~\ref{tab:DistinguishedLimit} also lists two new scaling regimes 
for these parameters that we suggest for use in subsequent asymptotic analyses
of (warm) moist air flow processes. The first regime, labelled $\alpha = 0$,
is consistent with similarity theory in the sense that we \emph{first} 
identify a set of dimensionless parameters for the system at hand and 
\emph{then} introduce a coupled limit between these parameters to enable 
asymptotic analyses. This limit is shown to (i) embed the asymptotics of moist 
air systematically within the general modelling framework reported upon in 
\cite{Klein2010}, and to (ii) constitute a ``rich limit'' in the sense that a 
maximum of the effects of water vapor on the equation of state of moist air are 
maintained at leading and first order. A somewhat awkward feature of this limit 
is, however, that the scalings in terms of the small parameter $\e$ do not match 
well in all cases with the actual numbers the dimensionless parameters take for 
moist air.

The second regime, in contrast, has been defined purely on the basis
of the actual magnitudes of the dimensionless parameters. Numbers between 
$0.4$ and $3.0$ are considered of order unity, while smaller or larger values
are associated with asymptotic rescalings in terms of $\e$. This provides a
scaling that better matches with the actual numbers than the first regime, 
but it is not strictly consistent with similarity theory. Although this is 
at odds with the usual procedures, it may actually open up an interesting route
of investigation. The thermodynamics of moist air may just be asymptotically 
compatible with a family of equation systems that features the same functional 
forms in the constitutive equations as those of moist air, but whose set of 
determining parameters is less constrained. The results of 
section~\ref{sec:MoistAdiabatAsymptoticsAndNumerics}, in which we compare 
asymptotic and error-controlled numerical approximations to the moist adiabatic 
distribution, corroborate this point of view.


\subsubsection{Reduced dynamical models for up- and downdrafts}

\paragraph{Updrafts in saturated air:}
Under the moist physics closure with refined thermodynamics and on the 
vertical advective time scale a dominant balance of forces is found in the 
vertical momentum equation. Positive buoyancy due to a potential temperature 
perturbation, $\tilde\theta$, which results from positive CAPE, is neutralized 
by the influence of the water vapor and rain water mixing ratios $q_v$ and $q_r$
onto the effective buoyancy force. In the precipitating core of a cloud tower we 
have, for the pragmatically defined distinguished limit for the cloud variables,
\beq\label{eq:VerticalBalance}
\tilde\theta 
+ \left(\frac{1}{E}-1\right)\left(\overline{q}_{vs} - \overline{q}_{v}\right)
-q_r
= 0\,,
\eeq
where $\overline{q}_{vs}(z)$ and $\overline{q}_{v}(z)$ are the saturation 
water vapor mixing ratio attained in the convective core and the mean water 
vapor content in the environment of the hot tower, respectively. These depend 
on the vertical coordinate, $z$, only. We recall that mixing ratios by definition 
compare the density of the gas component to the density of dry air. Moreover 
$E = R_d/R_v$ is the ratio of the dry 
air and vapor gas constants (see Table~\ref{tab:DistinguishedLimit}). Cloud water 
does not enter the buoyancy term to leading order, since it is one order of 
magnitude smaller than the other moisture components.

Next we note that $\tilde\theta$ and $q_r$ at the same time satisfy the transport 
equations
\beq
\label{eq:QRTransportSat}
\dss D_t {q}_{r} - \frac{1}{\overline{\r}}\pa_z(\overline{\r} V_r q_{r})
  & = 
    & \dss -w \, \frac{d \overline{q}_{vs}}{dz}  + \D_{q_r} 
      \\
\label{eq:ThetaTransportSat}
\dss D_t \tilde\theta   
  & = 
    & \dss w \frac{d\Delta\overline{\theta}}{dz}
          - k_l q_r (w-V_r) \frac{dT^{(1)}}{dz} + \D_{\theta} 
\eeq
where $D_t$ is the total time derivative, and 
$\overline{\rho}$ is the background density, 
$\Delta\overline{\theta}$ is the second-order difference between the moist
adiabatic and the background potential temperature stratifications, 
$V_r$ is the terminal droplet sedimentation velocity, 
$w$ is the vertical flow velocity, 
$k_l$ is a scaled ratio of the specific heats of liquid water and dry air, 
and $T^{(1)}$ is the first-order temperature stratification. The general
source terms $\D_{\theta}$ and $\D_{q_r}$ represent entrainment due to lateral 
turbulent transport and related effects. Neglecting these latter terms to focus just
on the vertical balances, and combining the balance from \eq{eq:VerticalBalance} 
and the transport equations in \eq{eq:QRTransportSat} one finds a diagnostic 
relation for the vertical velocity, 
\beq\label{eq:WEquation}
w \Biggl(k_l q_r \frac{dT^{(1)}}{dz} + \frac{d\Delta\overline{\theta}}{dz}
  & - 
    & \frac{1}{E} \frac{d \overline{q}_{vs}}{dz}
       + \left(\frac{1}{E}-1\right) \frac{d \overline{q}_{v}}{dz} 
      \Biggr)
      \nonumber\\ 
  & = 
    & k_l q_r V_r \frac{dT^{(1)}}{dz} 
    - \frac{1}{\overline{\r}} \pa_z(\overline{\r} V_r q_{r})
\eeq
%
which holds as long as $w > 0$, while the leading order vertical velocity is
zero otherwise within the precipitating core. Thus we have 
\beq
w = W(\pa_z q_r,q_r,z) := \max\left(0, A(q_r,z)\, q_r - B(q_r,z)\, \pa_z q_r\right)
\eeq
where $A(q_r,z), B(q_r,z)$ are straightforward abbreviations for terms from 
\eq{eq:WEquation}.

Inserting into \eq{eq:QRTransportSat} and neglecting horizontal transport in a 
quasi-onedimen\-sional approximation we find a Hamilton-Jacobi type equation for
the scaled precipitation mixing ratio,
\begin{equation}
\pa_t q_r 
=
W(\pa_z q_r,q_r,z) \left[\pa_z q_r + \frac{d\overline{q}_{vs}}{dz}\right] 
+ \frac{1}{\overline{\r}}\, \pa_z\left(\overline{\r} V_r q_r \right)
\,.
\end{equation}
Numerical solutions to this equation are discussed in 
section~\ref{sec:UpAndDownDrafts} below.

\paragraph{Downdrafts in undersaturated air:}
In undersaturated regions, all cloud water rapidly evaporates, so that its 
mixing ratio $\qc$ vanishes to leading order. Rain water, in contrast, 
evaporates at a rate of order unity on the convective time scale, so that 
the mixing ratios for vapor and rain, $\qv, \qr$, to leading order satisfy 
the transport equations 
\beq\label{eq:UnderSaturatedTowerDynamics}
\begin{array}{rcl}
\dss D_t \qv  
  & = 
    & \dss S_{ev} 
      \\[0pt]
\dss D_t \qr - \frac{1}{\r}  \pa_z(\r V_r \qr)  
  & = 
    & \dss - S_{ev}
\end{array}
\qquad\textnormal{with}\qquad
S_{ev} = L  C_{ev} (\qvs-\qv)q_{r}\,.
\eeq
The stability associated near moist adiabatic stratification is overcome 
in undersaturated regions only by evaporative cooling. The vertical velocity 
then follows the ``weak temperature gradient approximation''  \cite{HeldHoskins1985,SobelEtAl2001,KleinMajda2006}, 
\beq\label{eq:WTG}
w^{(0)}\frac{d\theta^{(1)}}{dz}=-L S_{ev}^{(0)}\,.
\eeq 
Eqs \eq{eq:UnderSaturatedTowerDynamics} and \eq{eq:WTG}, again in the quasi-1D
approximation, constitute the tower model for undersaturated air.
Numerical solutions to these undersaturated tower equations are also 
discussed in section~\ref{sec:UpAndDownDrafts} below.


\section{Governing equations}
\label{sec:GoverningEquations}

To describe the flow of moist air in the atmosphere we adopt the compressible
flow equations with a bulk microphysics closure scheme borrowing from 
\cite{Ooyama2001,Bannon2002,CottonEtAl2011}. We work with the dry air 
mass-averaged velocity, $\bfv$, such that the dry air mass balance reads
\beq\label{dry}
\pa_t\r_d + \nabla\cdot(\r_d  \bfv)=0\,,
\eeq
where $\rho_d$ is the dry air mass density. Of the three moisture components, 
vapor, cloud water, and rain water, we assume the former two to be advected 
by the dry air flow velocity, $\bfv$, whereas the rain water component 
falls at the terminal sedimentation velocity $V_r$. 

Based on these preliminaries, the total mass, momentum, potential temperature,
and moisture species balances read
\beq
\label{eq:MoistMass}
\pa_t\r_d + \nabla\cdot(\r_d \bfv ) 
  & = 
    & 0
      \\
\label{eq:MoistHorMom}
D_t {\bfu} +(2{\bfOmega}\times {\bfv})_{\|} + \frac{1}{\r}\nabla_{\|}p
  & =
    & \frac{\r_d}{\rho} q_r V_r \pa_z \bfu + {{\mathcal{D}}_{{\bfu}} }
      \\
\label{eq:MoistVertMom}
D_t  w + (2{\bfOmega}\times {\bfv})_{\bot} + \frac{1}{\r}\pa_z p
  & = 
    & - g + \frac{\r_d}{\rho} q_r V_r \pa_z w + \mathcal{D}_w 
      \\
\label{eq:MoistPotTemp}
C D_t\ln \theta + R D_t\ln p + \frac{L(T)}{T} D_t q_v
  & = 
    & c_l q_r V_r \left(\pa_z \ln \theta + \frac{R_d}{c_{pd}} \pa_z \ln p \right)
      + \mathcal{D}_\theta \quad
      \\
D_t q_v 
  & =
    & S_{ev} - S_{cd}+\mathcal{D}_{q_v}
      \\[5pt]
D_t q_c 
  & =
    & S_{cd} - S_{ac}- S_{cr}+\mathcal{D}_{q_c}
      \\
D_t q_r - \frac{1}{\r_d}\pa_z(\r_d q_r V_r) 
  & =
    & S_{ac} + S_{cr}- S_{ev}+\mathcal{D}_{q_r}
\eeq
where
\beq
\r 
  & = 
    & \r_d \, (1 + q_v + q_c + q_r)
      \\[5pt]
p 
  & = 
    & \r_dR_dT\left(1+\frac{q_v}{E}\right)
      \\[5pt]
T 
  & = 
    & \theta \left(\frac{p}{\rfr{p}}\right)^{\frac{\gamma-1}{\gamma}}
      \\[5pt]
\bfv 
  & = 
    & \bfu + w \bfk 
      \\[5pt]
C 
  & = 
    & c_{pd}+c_{pv}q_v + c_l (q_c + q_r) 
      \\[5pt]
R 
  & = 
    &  \Big(\frac{c_{pv}}{c_{pd}} R_d - R_v \Big)q_v + \frac{c_l}{c_{pd}} R_d (q_c + q_r) 
\eeq
In these equations $(\r, T, \theta, p, {\bfu}, w, q_v, q_c, q_r)$ are the density, 
temperature, potential temperature, pressure, the horizontal and vertical velocity 
components, and the mixing ratios of water vapor, cloud water, and rain water, 
respectively, $g$ is the gravitational acceleration, $\mathbf{\Omega}$ denotes the 
Earth rotation vector, and the 
subscripts ${ }_\bot$ and ${ }_{\|}$ refer to vertical and horizontal 
components respectively.  The turbulent and molecular transport terms are 
indicated by ${\mathcal D}$'s. Furthermore, $(c_{pd}, c_{pv})$ 
are the specific heat capacities at constant pressure of dry air and water vapor, 
$c_l$ is the heat capacity of liquid water, $(R_d, R_v)$ are the dry air and water 
vapor gas constants, $\gamma = c_{pd} / (c_{pd} - R_d)$ is the isentropic exponent 
of dry air, $\bfk$ is the vertical unit vector, $\rfr{p} = \unit{10^5}{\pascal}$ 
is a reference pressure,  and the Lagrangian time derivative is 
\beq
 D_t
  & =
    & \pa_t + {\bfv}\cdot \nabla = \pa_t + {\bfu}\cdot \nabla_{\|} + w \pa_z\,.
\eeq

In this work we assume constant heat capacity, $c_{pd}$ and $c_l$, of dry air
and liquid water, which implies that the latent heat of condensation $L$ is 
linear in the temperature
\beq\label{eq:LatentHeatFctn}
L(T)
=
L_{\textnormal{ref}} + (c_{pv}-c_l)(T-T_{\textnormal{ref}})
\equiv \rfr{L}\phi(T) \,.
\eeq

The source terms $S_{ev}, S_{cd}, S_{ac}, S_{cr}$ are the rates of evaporation 
of rain water, the condensation of water vapor to cloud water and the inverse 
evaporation process, the auto-conversion of cloud water into rainwater by 
accumulation of microscopic droplets, and the collection of cloud water by 
falling rain. To close the moisture dynamics we adopt the setting of 
\cite{KleinMajda2006} corresponding to a basic form of the bulk microphysics 
closure in the spirit of Kessler \cite{Kessler1969} and Grabowski and 
Smolarkiewicz \cite{GrabowskiSmolarkiewicz1996}:
\beq
S_{cd}&=&C_{cd}(q_v-  q_{vs})q_c+C_{cn}(  q_v- q_{vs})^+q_{cn}\,,\\
S_{ev}&=&C_{ev}\frac{p}{\r}(  q_{vs}- q_{v})^+   q_r\,,\\
S_{cr}&=&C_{cr} q_c  q_r\,,\\
S_{ac}&=&C_{ac} (q_c-  q_{ac})^+\,,
\eeq
where $y^+ \equiv \max(0,y)$. 
Here $(C_{cd},C_{ev},C_{cr},C_{ac}, C_{cn})$ are rate constants, $q_{cn}$ quantifies the
presence of condensation nuclei, and $q_{ac}$ is a threshold for cloud water mixing ratio beyond which autoconversion of cloud water into precipitation becomes active. 
Note that  for cloudless air $(q_c=0)$ we have positive condensation on cloud nuclei in oversaturated areas, whereas the inverse evaporation is suppressed in undersaturated air. However, as we shall see below, the condensation term will be defined implicitly from the asymptotics through the equation of water vapor at saturation, which also corresponds to a common definition of the condensation source term in the literature, see e.g. also \cite{GrabowskiSmolarkiewicz1996,Thuburn2017}.  

The saturation threshold is given by the saturation vapor mixing ratio
\beq
q_{vs}=\frac{\r_{vs}}{\r_d}\,,
\eeq
which can be expressed in terms of the saturation vapor pressure $\es$ through
\beq\label{qvs}
q_{vs}(T,p)=\frac{E \es(T)}{p-\es(T)},\qquad \textnormal{where} \quad E=\frac{R_d}{R_v}\,.
\eeq
The saturation pressure follows the Clausius-Clapeyron relation,
\cite{CottonEtAl2011,Emanuel1994},
\beq\label{eq:ClausiusClapeyron}
\frac{d\ln \es}{dT}=\frac{\rfr{L}\phi(T)}{R_vT^2} \,.
\eeq

We do not take into account different temperatures for the liquid water, but 
note that the temperature of  large droplets with a diameter of 
$\sim \unit{1}{\centi\meter}$ might differ from the one of the surrounding air. 
For a  further discussion on the incorporation of a different temperature for 
liquid water droplets into the dynamics we refer to \cite{Bannon2002}.


\section{Non-dimensionalisation and asymptotic scalings}
\label{sec:Nondimensionalisation}


\subsection{Dimensionless characteristic quantities}
\label{ssec:DimensionlessQuantities}

Standard thermodynamic reference values for the tropics are 
\beq\label{eq:ReferencesThermodynamics}
p_{\textnormal{ref}}       = 10^5 \ \textnormal{Pa}\,, 
\qquad T_\textnormal{ref} = 300\ \textnormal{K}\,, \qquad \r_{\textnormal{ref}}=\frac{p_\textnormal{ref}}{R_dT_\textnormal{ref}}\approx 1.16\ \textnormal{kg/m}^3 .
\eeq
To address deep convection phenomena it should be appropriate to use the ``bulk 
convective scales'' for space and time, \ie, the pressure scale height for 
horizontal and vertical length scales and the associated advection time scale 
based on a typical wind speed $\rfr{u}$,
\beq\label{eq:ReferencesSpaceTime}
\rfr{\ell} = h_{\textnormal{sc}} = \frac{p_\textnormal{ref}}{g \r_\textnormal{ref}}\approx 8.8  \,\textnormal{km}\,,
\quad
\rfr{u} \approx \unit{10}{\meter\per\second}\,,
\quad
\rfr{t} = \frac{\rfr{\ell}}{\rfr{u}} \approx \unit{15}{\minute} \,,
\eeq
where the gravitational acceleration is  $g=9.81\ \textnormal{m/s}^{2}$. 
A velocity of about $\unit{10}{\meter\per\second}$ corresponds to thermal wind 
scaling based on the global equator-to-pole potential temperature difference,
$\Delta\theta \sim \unit{30 \dots 49}{\kelvin}$, and thus constitutes a good 
generic flow speed reference value, see \cite{Klein2010} for further discussion.
Note that $\Delta\theta$ is also characteristic for the vertical variation 
of potential temperature across the troposphere and this will be crucial in 
the sequel. 

Based on these reference quantities the principal fluid dynamical parameters, 
\ie, the Mach, barotropic Froude, and bulk convective scale Rossby numbers are
\beq
\textnormal{M} 
= \frac{u_{\textnormal{ref}}}{\sqrt{p_{\textnormal{ref}}/\r_{\textnormal{ref}}}}
=\textnormal{Fr}
=\frac{u_{\textnormal{ref}}}{\sqrt{g h_{\textnormal{sc}}}}
\approx \frac{1}{30}, 
\qquad 
\textnormal{Ro}_B = \frac{u_{\textnormal{ref}}}{|{2 \bfOmega}_{\|}| h_{\textnormal{sc}}} \approx 10\,.
\eeq

The thermodynamics of moist air is characterized by several further dimensionless
quantities. At the standard temperature of $T_0 = \unit{273.15}{\kelvin}$ the specific 
heat capacities and gas constants and the latent heat of condensation amount to the 
values given in table~\ref{tab:ThermodynamicsParameters}.The heat capacities vary slightly 
with temperature, but these variations are small enough to not affect the expansions 
carried out below and they are therefore neglected in the following. The temperature
dependence of the latent heat is considerable, however, and it has been accounted for
already in \eq{eq:LatentHeatFctn}. Using the latter, we see that the reference value $\rfr{L}=L(\rfr{T})$ for the tropical reference value $\rfr{T}=300$~K only deviates slightly with $\rfr{L}\approx 2.44 \cdot 10^6$ J/kg.  The reference value $\rfr{L}/(c_{pd}\rfr{T})$ then reduces to approximately $8.1$, such that the stated orders of magnitudes in table~\ref{tab:DistinguishedLimit} remain unchanged even for the tropical conditions. The listed parameters  give rise to five 
independent dimensionless characteristic ratios and some more derived quantities
as listed in table~\ref{tab:DistinguishedLimit}. 

The mixing ratios of the water constitutents are dimensionless by 
definition, and their typical magnitude in the atmosphere is set by the saturation 
water vapor mixing ratio at reference conditions, \cite{Allaby2001},
\beq\label{qvs.ref}
q_{vs,\textnormal{ref}}=q_{vs}(p_{\textnormal{ref}},T_{\textnormal{ref}})\approx 0.022\,.
\eeq
%


\subsection{Distinguished asymptotic limits} 
\label{ssec:DistinguishedLimits}

As explained, e.g., in \cite{Klein2010}, asymptotic analysis in the presence of 
multiple small parameters generally requires the introduction of distinguished 
limits to uniquely identify one of many possible asymptotic limit regimes. The 
cited review introduced a particular distinguished limit for atmospheric modelling 
that couples the Mach, Froude, and Rossby numbers, and that has turned out to be 
rather uniformly useful across many different applications of scale analysis and 
asymptotics for the atmosphere. Specifically, this limit amounts to letting
\begin{equation}
\textnormal{M} = \textnormal{Fr} = \bigoh{\e^\frac{3}{2}}\,,
\qquad \textnormal{Fr}_{\textnormal{int}} = \bigoh{\e}\,,
\qquad \textnormal{Ro}_{\textnormal{B}} = \bigoh{\eps^{-1}}\,,
\end{equation}
where $\textnormal{Fr}_{\textnormal{int}} = \rfr{u}/c_{\textnormal{int}}$ is 
the Froude number based on a typical internal wave speed 
$c_{\textnormal{int}} = \sqrt{gh_{\textnormal{sc}}} \sqrt{\Delta\theta/\rfr{T}}$.
Since, therefore,  
$\textnormal{Fr}_{\textnormal{int}} = \textnormal{Fr}/\sqrt{\Delta\theta/\rfr{T}}$,
this leads to
\begin{equation}\label{eq:DelThetaMag}
\frac{\Delta\theta}{\rfr{T}} = \bigoh{\e}\,,
\end{equation}
where $\Delta\theta \approx \unit{30 ... 40}{\kelvin}$ measures the increase of
potential temperature across the height of the trophosphere. 
This sets a realistic magnitude of $\e$ to 
\beq\label{eq:EpsMag}
\e \sim 1/10\,.
\eeq

In the present paper, this distinguished limit for the dynamically relevant
parameters will be tied in with two alternative scaling regimes characterizing 
the moist thermodynamics of air. As in \cite{KleinMajda2006} we use \eq{qvs.ref} 
and \eq{eq:EpsMag} to let
\begin{equation}\label{eq:QvsMag}
q_{vs,\textnormal{ref}} = \bigoh{\e^2}\,.
\end{equation}
In agreement with the common assessment 
that the stratification of the near-equatorial troposphere is close to 
moist adiabatic, we identify the potential temperature difference 
$\Delta\theta$ with the typical temperature change associated with the total 
latent heat of condensation at saturation, \ie, $\Delta\theta \sim q_{vs} L/ c_{pd}$. 
Then, with \eq{eq:DelThetaMag} and \eq{eq:QvsMag} and evaluating at reference 
conditions we have
\begin{equation}
\frac{\Delta\theta}{\rfr{T}} = q_{vs,\textnormal{ref}} \frac{\rfr{L}}{c_{pd}\rfr{T}}
\quad\textnormal{\ie}\quad
\frac{\rfr{L}}{c_{pd}\rfr{T}} = \frac{L}{\e}
\quad\textnormal{with}\quad
L=\bigoh{1} \ \ (\e \to 0)\,.
\end{equation}
As in \cite{KleinMajda2006} we adopt the Newtonian limit for dry air, \ie,
\beq\label{new.lim}
\frac{R_d}{c_{pd}} = \frac{\gamma -1 }{\gamma}=\e \Gamma\,.
\eeq

Two alternative suggestions for coupled limit relations for the remaining 
characteristics of the moist thermodynamics of air are summarized in 
table~\ref{tab:DistinguishedLimit}. The scaling labelled $\alpha = 1$
in the right-most column pragmatically focuses on the
bare magnitude of the dimensionless ratios and assigns the respective 
asymptotic scalings accordingly. Consistently, numbers in the range
$0.4 \dots 3.0$ are considered of order $\bigoh{1}$, whereas numbers that
are substantially larger or smaller are assumed to scale with appropriate
powers of $\e$. While these assignments are consistent with the given 
magnitudes for the moist air parameters, they have the caveat of not 
representing a realizeable limit for a mixture of gases: Obviously, we obtain 
two expressions for, say, the ratio $R_d/R_v$, namely $E = \bigoh{1}$ 
and $(R_d/c_{pd}) (c_{pd}/R_v) = \e\Gamma A = \bigoh{\e}$ as $\e \to 0$.
Although awkward at first, this may not pose a major difficulty. The 
thermodynamics of moist air may just be asymptotically compatible with 
a family of equation systems that features the same functional forms in the
constitutive equations, but whose set of the determining parameters
is less constrained. The results of 
section~\ref{sec:MoistAdiabatAsymptoticsAndNumerics} below corroborate this 
point of view. 

In contrast, the scaling labelled $\alpha = 0$ appears to violate some 
basic order-of-magnitude estimates. Yet, this regime \emph{is} consistent 
with the thermodynamics of a mixture of gases in the sense of similarity 
theory, and each of its scalings as given in table~\ref{tab:DistinguishedLimit} 
is argued for in section~\ref{sec:MoistAdiabaticScalings} based on an analysis 
of the moist adiabatic hydrostatic state.  

Although the Coriolis force does not play a dominant role in the present flow regime, 
it nevertheless appears in the hot tower asymptotics below. At latitude $\varphi$, 
we decompose the earth rotation vector into vertical and horizontal components, 
\beq
2\bfOmega
= 2 \left(\bfOmega_\perp + \bfOmega_\|\right) 
= 2\Omega \sin(\varphi) \bfk + 2\Omega \cos(\varphi) \bfe_N
\eeq
where $\Omega = |\bfOmega|$. Over the small scales considered here, the latitude 
varies by $\bigoh{\e^3}$ (see \cite{Klein2010}), so that we can work with a constant
reference latitude $\varphi_0$. Interested in near equatorial flows, we 
consider $\varphi_0 = \bigoh{\e}$. This yields estimates for the Rossby numbers 
associated with the horizontal and vertical components of $\bfOmega$, 
\beq
\frac{2\bfOmega \hsc}{\rfr{u}} = \e^2 f \bfk + \e f_\perp \bfe_N + \littleoh{\e^2}
\eeq
where 
\beq
f = \frac{\sin(\varphi_0)}{\e^2 \Ro_B} = \bigoh{1} \, , \qquad 
f_\perp = \frac{\cos(\varphi_0)}{\e \Ro_B}  = \bigoh{1} \ 
\eeq
are the effective Coriolis parameters. With these approximations and in our
dimensionless notation, the Coriolis term reduces to
\begin{eqnarray}
\frac{\hsc}{\rfr{u}}\, \mathbf{2\Omega\times \mathbf{v}} =
\left(
\begin{array}{c}
- \e^2 f u + \e f_{\bot}w
  \\
\e^2 f v
  \\
- \e f_{\bot} u
\end{array}
\right)
=
\e \left(
\begin{array}{c}f_{\bot}w
  \\
0
  \\
-f_{\bot} u
\end{array}
\right) + O(\e^2)\,.
\end{eqnarray}
Typically the vertical Coriolis parameter $f_{\bot}$ is neglected but 
the term $\e f_{\bot}w$ in the horizontal momentum equation does contribute 
to the dynamics when the horizontal velocities in the cloud tower are 
$\bigoh{\e}$ and the characteristic time scale is that of advection in the 
vertical updraft (see \cite{CarqueEtAl2008} and the detailed analysis below). 


\subsection{Asymptotically rescaled governing equations} 
\label{sec.as.gov}

After non-dimensionalization, introduction of the coupled limits explained
in the previous section, and switching between the two alternative 
scalings from table~\ref{tab:DistinguishedLimit}, the governing equations
read
\beq
\label{as.res.dry}
\pa_t\r_d + \nabla\cdot(\r_d \bfv ) 
  & = 
    & 0
      \\
D_t {\bfu} + \e \bfF_{\|} + \frac{1}{\e^3}\frac{1}{\r}\nabla_{\|}p
  & =
    & \e^2 \frac{\r_d}{\rho} q_r V_r \pa_z \bfu 
      \\
\label{eq:MoistVertMomDimless}
D_t  w + \e \bfF_{\perp} + \frac{1}{\e^3}\frac{1}{\r}\pa_z p
  & = 
    & - \frac{1}{\e^3} + \e^2 \frac{\r_d}{\rho} q_r V_r \pa_z w 
      \\
\label{eq:MoistPotTempDimless}
C_\e  D_t\ln \theta + \e^2 R_\e D_t\ln p + \e \frac{L\phi_\e}{T} D_t q_v
  & = 
    & \e k_l q_r V_r \left(\pa_z \ln \theta + \e\Gamma \pa_z \ln p \right)
      \\[5pt]
\label{qv.res.or}
D_t q_v 
  & =
    & S_{ev} - S_{cd} 
      \\[5pt]
D_t q_c 
  & =
    & - \frac{1}{\e}S_{cr} + S_{cd} - S_{ac} 
      \\
\label{qr.res.or}
D_t q_r - \frac{1}{\r_{d}}\pa_z(\r_{d} q_r V_r)
  & =
    &  \frac{1}{\e} S_{cr} + S_{ac} - S_{ev}\ \ \ 
\eeq
where
\beq
\r 
  & = 
    & \r_d \, (1 + \e^2 \left[q_v + q_c + q_r\right])
      \\[5pt]
p
  & = 
    & \r_d T (1 + \e^{1+\alpha}\qv/E)
      \\[5pt]
\label{eq:ExnerDefinition}
T 
  & = 
    & \theta\, p^{\e\Gamma} \equiv \theta\, \pi
      \\[5pt]
\bfv 
  & = 
    & \bfu + w \bfk 
      \\[5pt]
\label{eq:Ce}
C_\e 
  & = 
    & 1  +  \e \left[\e^\alpha k_v q_v + k_l (q_c + q_r)\right] 
      \\[5pt]
\label{eq:Re}
R_\e
  & = 
    &  \Gamma k_l (q_c + q_r) + \e^\alpha \kappa_v q_v 
      \\[0pt]
\label{eq:LatentHeatTempDependence}
\phi_\e 
  & = 
    &  1 - \chi (T-1)\,, \qquad \left(\chi = \frac{k_l - \e^\alpha k_v}{L}\right)
\eeq
Here we neglected the turbulent and molecular transport terms, whose incorporation 
is left for future work, and we have introduced the Exner pressure~$\pi$ in 
\eq{eq:ExnerDefinition}. Since the \emph{horizontal} momentum balance will only be 
expanded to first order, the additional horizontal momentum contributions due to 
water loading  will not play a role in the leading order dynamics. This is in line 
with the common assumption that their contributions are of lesser importance~\cite{CottonEtAl2011}.

As in \cite{KleinMajda2006} we let 
\beq
\label{Cd.hat}
S_{cd}
  & = 
    &\frac{1}{\e^n}(C_{cd}(q_v-q_{vs})q_c+C_{cn}(q_v-q_{vs})^+q_{cn})\,,
      \\
\label{Cev.hat} S_{ev}
  & =
    &C_{ev} \frac{p}{\rho} (q_{vs}-q_{v})^+ q_r\,,
      \\
S_{cr}
  & = 
    &C_{cr} q_c q_r\,,
      \\
\label{Sac.hat}
S_{ac}
  & = 
    & C_{ac} (q_c- \e q_{ac})^+\,,
\eeq
where now all appearing rate constants are $O(1)$. 
As already mentioned above, the condensation term will be obtained from \eq{qv.res.or} 
at saturated conditions leading to $S_{cd}\approx-w\frac{dq_{vs}}{dz}$. Depending
on the scaling regime considered, the saturation mixing ratio is given as a function 
of pressure and temperature by
\beq\label{qvs.es}
q_{vs}(p,T) = \frac{E \es(T)}{p-\e^{1+\alpha} \es(T)}\,,
\eeq
and the asymptotically rescaled  Clausius-Clapeyron equation for the saturation vapor 
pressure, $\es$, reads 
\beq\label{eq:ClausiusClapeyronScaled}
\frac{d\ln \es}{dT} = \frac{LA}{\e} \frac{\phi_\e(T)}{T^2}\,
\eeq
with $\phi_\e(T)$ from \eq{eq:LatentHeatTempDependence}. 


\section{The moist adiabat and the asymptotic scalings for the moisture parameters}
\label{sec:RevisedScalings}

The present section justifies the rationale behind the distinguished limit 
for the moisture quantities labelled $\alpha = 0$ in table~\ref{tab:DistinguishedLimit}
by showing that the resulting moist adiabatic stratification is compatible with the
general asymptotics framework from \cite{Klein2010} as desired. This regime also 
turns out to be a ``rich limit'' in that it retains a maximal number of terms in the
equations governing the moist adiabat among a broader family of possible scalings. 

The leading and first order asymptotic solutions for the moist adiabat are 
obtained for both scaling regimes from table~\ref{tab:DistinguishedLimit}. Both turn 
out to compare favorably with a high-accuracy numerical solution based on the 
unapproximated equations.
  

\subsection{Consequences of different scaling limits for the moist adiabat}
\label{sec:MoistAdiabaticScalings}

The embedding of moist thermodynamics into the asymptotic modelling framework
requires a careful balance of various large and small quantities. This is 
to be achieved through the appropriate choice of a distinguished limit tying
these various quantities to one small reference parameter, $\e \ll 1$. To 
investigate the consequences which different choices of distinguished limits
would imply, we consider the following rather general scaling scheme  
\beq\label{eq:ThermodynamicScalingsGen}
\frac{\rfr{L}}{c_{pd}\rfr{T}} = \frac{L}{\e^a}\,,
\quad
\frac{R_d}{c_{pd}} = \e^{b}\Gamma\,,
\quad
\frac{R_v}{c_{pv}} = \e^{b_v}\Gamma_v\,, 
\quad
\frac{c_l}{c_{pd}} = \frac{k_l}{\e^{b_l}}\,,
\quad
\frac{R_d}{R_v} = \e^{c} E \,. \ \  
\eeq
Here $L, \Gamma, \Gamma_v$, and $E$ are $\bigoh{1}$ as $\eps\to 0$ and 
the (positive) exponents $a,b,b_v, b_l,c$ are not further specified as yet.
For a scaling regime appropriate for atmospheric applications, and consistent
with the developments of the previous section, we will discuss the consequences 
of chosing $a = b = 1$, and $b_v, b_l, c \in \{0,1\}$ below. For the 
scaling of the r.h.s.\ of the Clausius-Clapeyron relation in 
\eq{eq:ClausiusClapeyron} the choices in \eq{eq:ThermodynamicScalingsGen} imply 
\beq\label{eq:ActivationEnergy}
\frac{\rfr{L}}{R_v\rfr{T}} 
= \frac{1}{\e^{a+b-c}}\frac{LE}{\Gamma} 
\equiv \frac{AL}{\e^{d}} 
\qquad\textnormal{where}\qquad
d = a + b - c\,.
\eeq
Also we account for the generally small values of the saturation
water vapor mixing ratio $q_{vs}$ and pressure $\es$ by letting
their bare values before division by the reference value from 
\eq{qvs.ref} satisfy the scaling
\beq
\qvshat = \e^e\, \qvs \,, \qquad  \eshat = \e^{e-c}\, \es\,,
\eeq 
with $\qvs, \es = \bigoh{1}$ as $\e\to 0$. This is motivated by the 
constitutive relation between the saturation mixing ratio $\qvshat$ 
and the saturation water vapor pressure $\eshat$, 
\beq\label{eq:QvsEsScaling}
\e^e \qvs
= \qvshat 
= \frac{R_d}{R_v}\frac{\eshat}{p - \eshat}
= \frac{\e^e E \es}{p - \e^{e-c}\es}\,,
\eeq
where we used the last entry of \eq{eq:ThermodynamicScalingsGen}.

The equations governing the moist adiabatic stratification consist 
of the potential temperature evolution equation \eq{eq:MoistPotTempDimless}
specialized to a vertical column neglecting the dissipative source 
and sedimentation of rain terms and requiring exact balance of all 
vertical advection terms, and the vertical momentum equation 
\eq{eq:MoistVertMomDimless} specialized to hydrostatic balance.
With the scalings introduced above, and using the definition of the Exner 
pressure $\pi$ in \eq{eq:ExnerDefinition}, these equations become 
\beq
\label{eq:DThetaDzMoistAdiabaticGen}
\frac{d\ln \theta}{dz}
  & =
    & - \e^{e-a} \frac{L\phi_\e(T)}{C_\e \pi\theta} \frac{d}{dz}\qvs(\theta,\pi;\e)
      - \e^{e-b}\frac{R_\e}{C_\e} \frac{1}{\Gamma} \frac{d\ln \pi}{dz}
      \\
\label{eq:DPiDzMoistAdiabaticGen}
\frac{d\ln\pi}{dz}
  & =
    & - \e^b \frac{\Gamma}{\pi\theta}  
        \frac{1 + \e^e \qvs}{1+\e^{e-c}\qvs/E}
\eeq
For the saturation water vapor mixing ratio, $q_{vs}$, the dependence on 
the unknowns $\theta, \pi$ follows from the definition of $\es$ in 
\eq{eq:QvsEsScaling} which depends on temperature only so that, 
with $T = \pi\theta$,
\beq
\qvs(\theta,\pi;\e) 
= \frac{E \es(\theta\pi;\e)}{\pi^{1/\e^b\Gamma} - \e^{e-c}\es(\theta\pi;\e)}\,,
\eeq
where the Clausius-Clapeyron relation for $\es$ reads as 
\beq\label{eq:ClausiusClapeyronScaled2}
\frac{1}{\es}\frac{d\es}{dT} = \frac{1}{\e^{d}} \frac{AL}{T^2}\phi_\e(T)
\eeq
with 
\beq\label{eq:PhiOfT}
\phi_\e(T) 
= 1 + \chi_\e (T-1)
\qquad\textnormal{where}\qquad
\chi_\e = \e^{a-b_l} \frac{k_l}{L}- \e^{a+b-b_v-c}\frac{k_v}{L}
\eeq
and $\e^{b-b_v-c}k_v = \cpv/\cpd = (\Rd/\cpd) (\Rv/\Rd) (\cpv/\Rv) 
= \e^{b-b_v-c} \Gamma / (E\Gamma_v)$.

The scalings and governing equations for the moist adiabatic stratification
summarized in \eq{eq:ThermodynamicScalingsGen}--\eq{eq:PhiOfT}
are combined in Appendix~\ref{app:ScaledMoistAdiabaticEquation} to eliminate
the derivative $d\qvs/dz$ from \eq{eq:DThetaDzMoistAdiabaticGen}, and this 
yields a scaled effective equation for the potential temperature. Keeping all
dominant contributions for $\e \to 0$ (see appendix~\ref{sapp:EffectiveMADEquation} 
for what ``dominant'' refers to precisely), this equation reads
\beq\label{eq:DThetaDzFinal}
\frac{d\theta}{dz} 
= \e^b \frac{\Gamma}{\pi}
  \left( 1 - \e^{a+d-e}
            \frac{I_\e + \e^{e-a-b} \frac{L\phi_\e}{\Gamma T} \qvs}%
                 {\frac{AL^2\phi^2_\e}{T^2} \qvs + \e^{a+d-e} I_\e}
      \right)\,,
\eeq
where $I_\e = 1 - \e^{e-c}\es/p$. 
To obtain a $\theta$-variation that is small of order $\bigoh{\e^b}$ as desired
here, we must require $a+d-e \geq 0$, assuming for the moment that the remaining
powers of $\e$ in \eq{eq:DThetaDzFinalApp} have positive exponents. This will
be verified below in hindsight. When $a+d-e > 0$, \eq{eq:DThetaDzFinalApp} 
combined with the equation for the Exner pressure from 
\eq{eq:DPiDzMoistAdiabaticGen} yields
\beq
\begin{array}{rcl}
\dss\frac{d\theta}{dz} 
  & = 
    & \dss\e^{b} \frac{\Gamma}{\pi} (1 + \littleoh{1})
      \\[10pt]
\dss\frac{d\pi}{dz} 
  & = 
    & \dss- \e^{b} \frac{\Gamma}{\theta}
\end{array}
\qquad\textnormal{and hence}\qquad
\frac{dT}{dz} = \frac{d\theta\pi}{dz} = \littleoh{\e^{b}}\,, 
\eeq
\ie, one finds temperatures with lesser variation across the troposphere
than that of the potential temperature, which is not realistic. 

In contrast, when $a+d-e = 0$, the second term in the bracket in 
\eq{eq:DThetaDzFinalApp} is of order unity and there will be vertical 
variations of the background temperature to order $\e^b$, comparable to 
those of $\pi$ and $\theta$. Thus, using \eq{eq:ActivationEnergy}, we let
\beq\label{eq:FirstRichChoice}
e = a+d = 2a + b - c\,.
\eeq
Consistency with the unified asymptotic modelling framework
summarized in~\cite{Klein2010} requires potential temperature 
stratifications of order $\bigoh{\e}$, \ie, we let $b = 1$. 
Considering that $\rfr{L}/c_{pd}\rfr{T} \approx 9.1$ according to 
table~\ref{tab:DistinguishedLimit}, and with 
$\e \sim 1/10$, the most reasonable choice for $a$ in 
\eq{eq:ThermodynamicScalingsGen} is $a = 1$ as well. Going back to \eq{eq:DThetaDzFinal}
we observe that the second term in the numerator on the r.h.s.\ 
will be negligible asymptotically if $e > a+b$, while we obtain a
classical ``rich limit'' that maintains all effects covered by
\eq{eq:DThetaDzFinal} simultaneously if we let $e = a+b = 2$. This  
corresponds to $q_{vs} = \bigoh{\e^2}$ in line with the earlier 
order of magnitude assessment in \eq{eq:QvsMag}. With this rich limit
adopted, \eq{eq:FirstRichChoice} implies $c = 1$. To summarize, we
let
\beq
a = b = c = 1\, \qquad\textnormal{and}\qquad e = 2\,.
\eeq

Admittedly, letting $c = 1$, \ie, $R_d/R_v \approx 0.62 = \e E$ may
appear a bit extreme, and $c = 0$ seems more reasonable. Yet, this would 
imply $e = 3$ following the arguments given above, \ie, it would imply that 
the mixing ratios of the water constituents are $\bigoh{\e^3}$. This, 
in turn, is not satisfactory either: First, the mixing ratios
of the water constituents have a typical magnitude of $1\%$, and this matches
with $\bigoh{\e^2}$ much better than with $\bigoh{\e^3}$ 
(see fig.\ \ref{fig:MoistAdiabat} below). Secondly, the buoyancy effects
of water loading are generally thought to be important, and 
this leads us to an estimate of their contribution to CAPE.
Integrated across the troposphere, water 
loadings of order $1\% \sim \e^2$ amount to potential energies of order 
$10^{-2} \cdot g\hsc \sim \unit{10^3}{\meter\squared\per\second\squared}$.
This is a realistic level of CAPE in many situations and one would aim for
a scaling of the water constitutent mixing ratios that allows for 
their buoyancy effects to participate in the related dynamics. With 
$q \sim \bigoh{\e^3}$, however, the influence of water loading induced
buoyancy would be restricted to rather tame situations with CAPE merely 
of order $\unit{10^2}{\meter\squared\per\second\squared}$. 

Notably, the choice of $b_v$, \ie, whether or not we assume a 
Newtonian limit for the water vapor, does not play a role up to this
point. The last entry in the column for $\alpha = 0$ in table~1 provides 
a guideline, however. Under the adopted scaling, $R_v / c_{pd} = \bigoh{1}$ 
and it has an actual value for moist air of $0.46$. At the same time, the 
Newtonian limit for dry air implies $R_d / c_{pd} = \bigoh{\e}$. Thus, 
the combination
$\frac{c_{pv}}{c_{pd}}\frac{R_d}{c_{pd}} - \frac{R_v}{c_{pd}}$ is
asymptotically equal to $-R_v/c_{pd}$, and hence negative, for $b_v < 1$, 
while it is asymptotically large for $b_v > b$. Only $b = b_v = 1$ will
allow for the adjustment to a finite positive value in the limit, and 
this is why we adopt the Newtonian limit for water vapor as well with
$R_v / c_{pv} = \e \Gamma_v$.   

Given that $b_v = 1$, the exponent $b_l$, which scales the liquid water heat 
capacity in \eq{eq:ThermodynamicScalingsGen}, should be larger or equal to 
unity, as the liquid water heat capacity exceeds that of water vapor by 
a factor larger than two. It is not reasonable, on the other hand, to let
$b_l > 1$, because this would imply that the temperature dependence of the
latent heat would dominate relative to its reference value, see~\eq{eq:PhiOfT},
and this is not realistic: The temperature induced relative variation of latent 
heat across the troposphere amounts to only $5\%$. Thus we chose $b_l = 1$ as
well and this completes the discussion of the scaling for $\alpha = 0$ from 
table~\ref{tab:DistinguishedLimit}.


\subsection{Asymptotic analysis vs.\ numerical computation of the moist adiabat}
\label{sec:MoistAdiabatAsymptoticsAndNumerics}

Here we compare high-accuracy numerical solutions for the moist adiabatic hydrostatic 
state, described in detail by \eq{eq:DThetaDzMoistAdiabaticGen} and 
\eq{eq:DPiDzMoistAdiabaticGen}, with leading and first order asymptotic solutions 
under the two scaling regimes from table~\ref{tab:DistinguishedLimit}.

The full, somewhat lengthy, detail of the asymptotic analysis is given in 
appendix~\ref{app:MADAsymptotics}. Here we just summarize the leading order 
analysis to provide an impression of how the calculations proceed.
Keeping only the dominant terms in the equations so as to streamline the
exposition for this chapter, we have
\beq
\label{eq:ThetaEqnHydroSimplified}
\frac{d\theta}{dz} 
  & = 
    & \dss - \e \frac{L}{\pi}\frac{d\qvs}{dz}
      \\
\label{eq:PiEqnHydroSimplified}
\frac{d\pi}{dz} 
  & = 
    & - \e \frac{\Gamma}{\theta}
\eeq
together with the constitutive relations
\beq\label{eq:EOSSimplified}
& \dss \qvs = \frac{E\es(T)}{p}\,,
\qquad
\dss p = \pi^{1/\e\Gamma}\, ,
\qquad
T = \pi\theta\,,
\qquad
\dss \frac{1}{\es}\frac{d\es}{dT} = \frac{1}{\e} \frac{AL}{T^2}\,.
\eeq
The unknowns $\theta, \pi$ are expanded as
\beq
\label{eq:HydrostaticExpansionScheme}
\begin{array}{rcl}
\dss \theta 
  & = 
    & \dss 1 + \e \theta^{(1)}(z) + \littleoh{\e}
      \\[5pt]
\dss \pi
  & = 
    & \dss 1 + \e \pi^{(1)}(z) + \littleoh{\e}
\end{array}\,.
\eeq
Inserting the expansion of $\theta$ into \eq{eq:PiEqnHydroSimplified} we obtain
\beq\label{eq:PiOneMADLeadingOrder}
\pi\order{1}(z) = -\Gamma z\,,
\eeq
and then the constitutive equations for pressure and temperature in 
\eq{eq:HydrostaticExpansionScheme} yield
\beq
\label{eq:THydroExpansion}
T = 1 + \e T\order{1} + \littleoh{\e}\, , 
\qquad 
T\order{1}(z) 
  & = 
    & \theta\order{1}(z) -\Gamma z
      \\
\label{eq:PHydroExpansion}
p = p\order{0} + \e p\order{1} + \littleoh{\e}\,,
\qquad
p\order{0}(z) 
  & =
    & \lim\limits_{\e\to 0} \left(1 - \e\Gamma z\right)^{\frac{1}{\e\Gamma}} = e^{-z}\,.
\eeq

Next we observe that $\qvs$ is given as a function of temperature and pressure, 
$(T,p)$, and these are functions of our primary unknowns $\theta, \pi$ in turn. 
After a short calculation applying the chain rule appropriately, we have 
\beq
\frac{d\qvs}{dz} 
  & = 
    & \qvs \left(\frac{d\ln\es}{dT} 
                 \left[\pi \frac{\pa\theta}{dz} + \theta \frac{\pa\pi}{dz}\right] 
           - \frac{d\ln p}{d\pi} \frac{d\pi}{dz}
     \right)
      \\
  & =
    & \qvs \left( \frac{AL}{T^2} \left[- L\frac{d\qvs}{dz} - \Gamma\right] + \frac{1}{\theta\pi}
      \right)\,.
\eeq
Using the asymptotic ansatz for $\theta,\pi,T,p$, keeping only the leading terms,
and solving for $(dq_{vs}^{(0)}/dz)^{-1}$, 
\beq
\frac{dz}{dq_{vs}^{(0)}} 
  & =
    & - \frac{1}{AL\Gamma - 1}\left(AL^2 + \frac{1}{q_{vs}^{(0)}}\right)\,.
\eeq
This is readily solved by 
\beq
\label{eq:QvsZeroMADLeadingOrder}
\ln \left(\frac{q_{vs}^{(0)}(z)}{{q_{vs}^{(0)}}_0}\right) 
+ AL^2\, \left(q_{vs}^{(0)}(z) - {q_{vs}^{(0)}}_0\right) = - (AL\Gamma - 1)\, z\, .
\eeq
Returning to \eq{eq:ThetaEqnHydroSimplified} and keeping again only the leading order
terms we find 
\beq\label{eq:ThetaOneMADLeadingOrder}
\theta^{(1)}(z) = -L \left(q_{vs}^{(0)}(z) - {q_{vs}^{(0)}}_0 \right) + \theta\order{1}_0\,,
\eeq
for the first-order potential temperature. Here 
$\theta\order{1}_0 \equiv \theta\order{1}(0)$ captures deviations of the 
near-surface temperature from the reference temperature used in the 
non-dimensionalization of the equations. 

Equations \eq{eq:PiOneMADLeadingOrder}, \eq{eq:QvsZeroMADLeadingOrder}, and 
\eq{eq:ThetaOneMADLeadingOrder} describe the moist adiabatic profile asymptotically
to leading order in $\e$. This leading-order solution is shared by both  
scalings from table~\ref{tab:DistinguishedLimit}. For comparison, numerical 
solutions of the full equations for the moist adiabat without asymptotic 
approximation were obtained using the MatLab {\tt ode23s()} routine. Variations 
of the error tolerance option of the routine in the range 
${\tt tol} = 10^{-4} \dots 10^{-10}$ produced indistinguishable results at the 
level of the graphics output.

\begin{figure}
  \centering
  \includegraphics[width=0.8\textwidth]{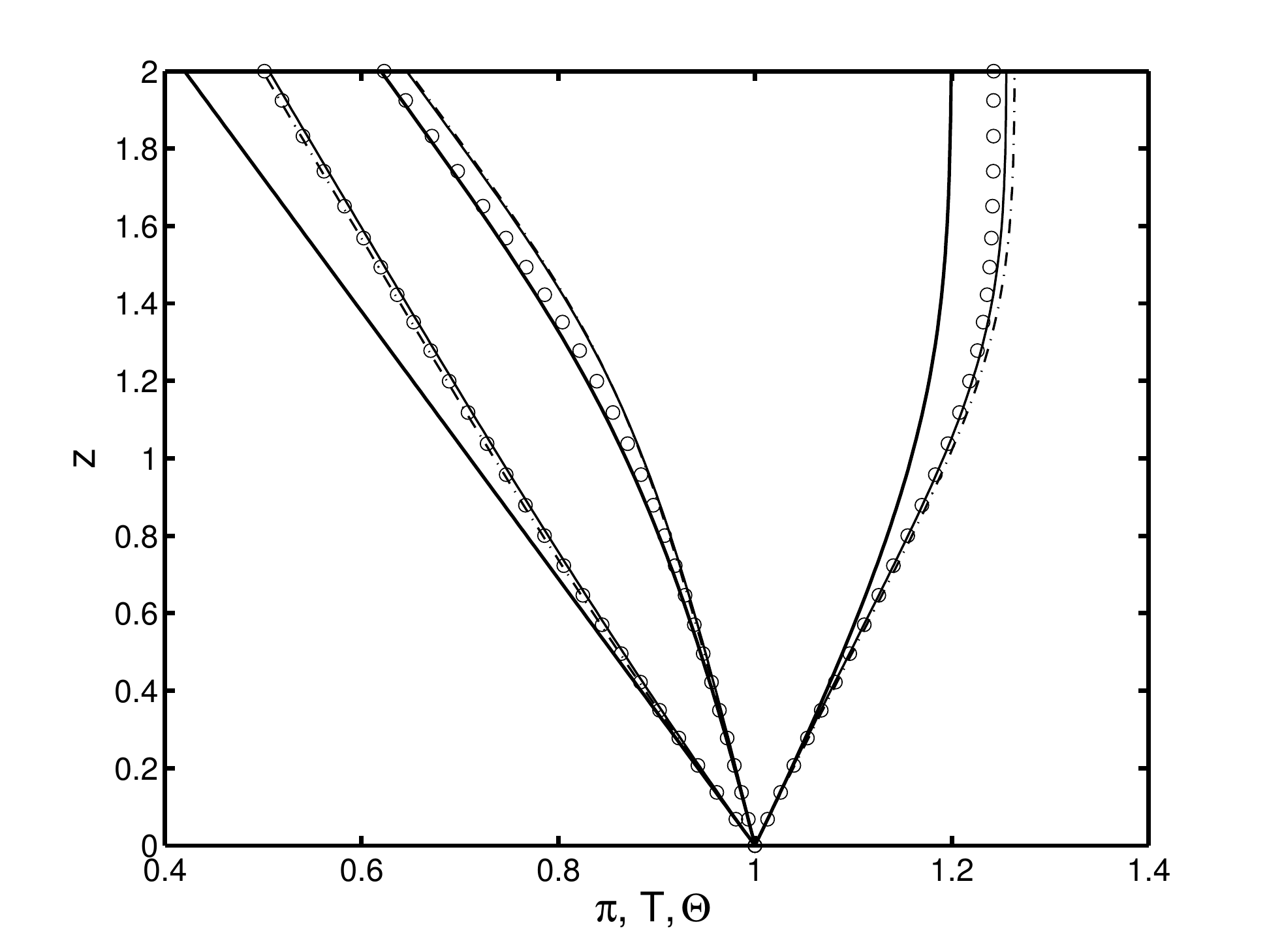}
\caption{Comparison of asymptotic and highly accurate numerical approximations to
the hydrostatic moist adiabat. thick solid lines: leading-order asymptotics; 
thin solid line(s): first-order accurate asymptotics for the regime with $\alpha = 0$; 
dash-dotted lines: first-order accurate asymptotics for $\alpha = 1$; circles: 
error-controlled numerical solution of the full moist adiabatic equations 
(only every third data point used by the adaptive MatLab routine {\tt ode23s()} is 
displayed). Leftmost triple of distributions: Exner pressure $\pi$; middle triple:
temperature $T$; right triple: potential temperature $\theta$. 
}
\label{fig:MoistAdiabat}       
\end{figure}

The thick solid lines in figure~\ref{fig:MoistAdiabat} represent the leading-order 
asymptotic solutions for Exner pressure $\pi$ (leftmost line), temperature 
$T = \theta\pi$ (line in the middle), and potential temperature $\theta$ 
(rightmost line). The open circles display the numerical solutions. There is 
good qualitative agreement, whereas quantitative accuracy leaves room for improvement. 

To improve on this, we worked out the next order asymptotic corrections in 
appendix~\ref{app:MADAsymptotics} for both scaling regimes and the results 
are included in fig~\ref{fig:MoistAdiabat} as well. The thin solid lines represent 
the asymptotic solutions to leading and first order for the scaling labelled 
$\alpha = 0$, whereas the thin dash-dotted lines represent the same for $\alpha = 1$. 
Deviations of these approximations from the numerical results are in the 
range of a few percent so that good quantitative accuracy is now obtained as well.

Although one might expect even better accuracy formally, since $\eps \sim 1/10$ 
and the left-over truncation errors are $\bigoh{\e^3}$, we consider the results
in fig.~\ref{fig:MoistAdiabat} to be quite satisfactory for the following reason:
The Newtonian approximation of the equation of state from \eq{new.lim} sets 
$(\gamma-1)/\gamma = \e \Gamma$ as $\e\to 0$, while the given concrete model 
parameters produce a value of $(\gamma-1)/\gamma = 0.29$, which is not a very
small number. This places limits on the accuracy that can be expected from the
lowest-order asymptotic approximations.
%


\section{Convective time scale dynamics of an upright cloud tower}
\label{sec:CloudTowerDynamics}

In this section we adopt the scaling regime with $\alpha = 1$ from 
table~\ref{tab:DistinguishedLimit}, discussing deviations between the two regimes 
briefly at the end in a separate subsection. 


\subsection{Cloud tower scaling}
\label{sec:CloudTowerScaling}

\begin{figure}[htbp]
\begin{center}
\includegraphics[height=4cm]{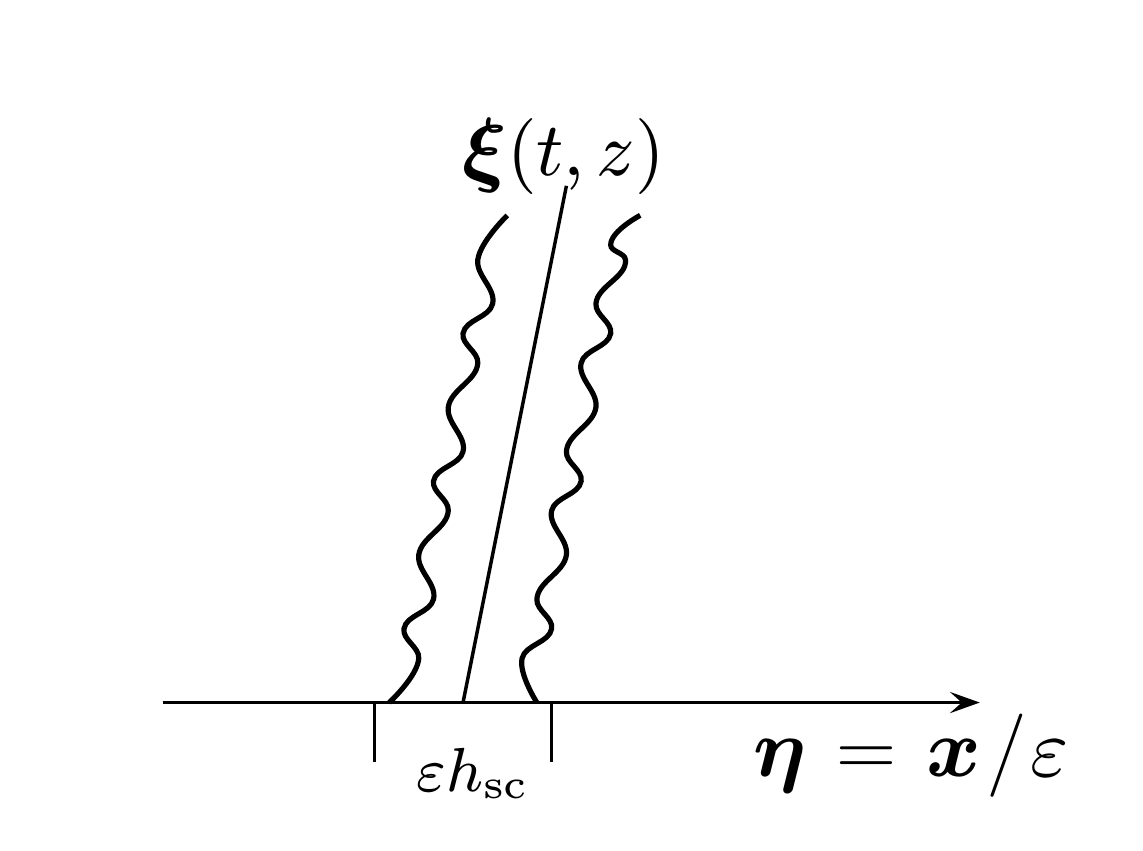}
\caption{Deep convective tower scaling}
\label{fig:CloudTowerSketch}
\end{center}
\end{figure}

Here we study flows within a deep convective cloud tower with vertical extent 
comparable to the pressure scale height, $\hsc \sim \unit{10}{\kilo\meter}$, but 
with narrow horizontal support of order $\e\hsc \sim \unit{1}{\kilo\meter}$, 
see fig.~\ref{fig:CloudTowerSketch}. We restrict to a single cloud tower embedded
in a quiescent environment to focus just on the dynamics of convection within. 
To resolve the small horizontal scale we introduce the stretched coordinate
\beq
 \bfeta =\frac{\bfx}{\e}\,, 
\eeq
as sketched in fig.~\ref{fig:CloudTowerSketch}, so that the solution ansatz to 
order $\bigoh{\e^N}$ for any of the unknowns, $\phi$, reads
\beq
\phi(t,\mathbf{x},z;\e)
  & = 
    & \sum_{i=0}^{N} \e^{i} \phi^{(i)}(t, \bfeta, z) + \littleoh{\e^N}\,,
      \\
\phi^{(i)}(t, \bfeta, z)
  & = 
    & \Phi_i(z) + \phitilde^{(i)}(t, \bfeta, z)
\eeq
\ie, we split the perturbation functions into their purely $z$-dependent 
background contributions and their small-scale variations within the cloud tower. 

We are interested in dominantly vertical updrafts with updraft velocities of
order unity, \ie, of order $\unit{10}{\meter\per\second}$ in dimensional terms,
developing on the $\bigoh{1}$ time scale, \ie, on time scales of order
$\hsc/\rfr{u} \sim \unit{20}{\minute}$ in dimensional terms. We anticipate 
that this implies either a constant or vanishing background wind and we assume, 
if necessary, a moving coordinate system within which the background state is
stationary. In this frame of reference the horizontal velocity is expanded as
\beq\label{ans.u}
\bfu=\e \bfu^{(1)}+\e^{2}\bfu^{(2)}+O(\e^3)\,,
\eeq 
whereas the vertical velocity component has a leading order contributions 
describing the intense up- and downdrafts of interest
\beq
w=w^{(0)}+\e w^{(1)}+O(\e^2)\,.
\eeq


\subsection{Rescaled governing equations}

The rescaled governing equations are obtained by replacing 
\beq
\nabla_{\|} 
  &\rightarrow 
    &  \frac{1}{\e}\nabla_\bfeta\,,
\eeq
leading, in particular, to the transport operator 
\beq\label{Dt}
D_t = \pa_t + \frac{1}{\e}\bfu\cdot \nabla_\bfeta + w\pa_z \,. \ \ 
\eeq
Then the scaled governing equations become
\beq\label{as.cont}
 \pa_t\rd + \frac{1}{\e}\nabla_\bfeta \cdot (\rd \bfu) + \pa_z(\rd w) 
   & = 
     & 0\,,
       \\
\label{as.u}
D_t\bfu + \e\mathbf{F}_{\|}
  & = 
    & - \frac{1}{\e^4}\frac{1}{\r}\nabla_\bfeta p + \e^2 \frac{\rd}{\r}\qr V_r \pa_z \bfu\,,
      \\
\label{as.w}
D_t w + \e\mathbf{F}_{\perp}
  & = 
    & -\frac{1}{\e^{3}}\frac{1}{\r}(\pa_z p  +\r ) + \e^2 \frac{\rd}{\r}\qr V_r \pa_z w\,,
      \\
\label{as.theta}
C_\e D_t \ln \theta + \e^2 R_\e D_t\ln p 
  & - 
    & \e k_l q_r V_r 
      \left(\pa_z \ln \theta  +  \e \Gamma \pa_z \ln p
      \right)
      \nonumber\\
  & =
    & \e\frac{L}{T} (1- \chi (T-1))(S_{cd}-S_{ev}) 
\eeq
with $C_\e, R_\e$ from \eq{eq:Ce} and \eq{eq:Re}, respectively. 
Accordingly, we have
\beq
\label{as.qv}D_t q_v&=&  S_{ev} -S_{cd} \,,\quad \\
\label{as.qc}D_t q_c&=&   S_{cd} - \frac{S_{cr}}{\e} - S_{ac}  \,, \quad \\
\label{as.qr}D_t q_r -\frac{1}{\r_d}\pa_z(\r_d q_rV_r) &=&  \frac{S_{cr}}{\e} - S_{ev} +   S_{ac} \,, \quad 
\eeq
for the moisture dynamics, and we conclude from the horizontal momentum balance 
in \eqref{as.u} that the expansion of the thermodynamic quantities about the 
hydrostatic background remains valid at least up to $O(\e^2)$
\beq
\r=\r_h+O(\e^2)\,,\qquad  p=p_h+O(\e^2)\,.
\eeq
%


\subsection{Small scale dynamics}
\label{ssec:SmallScaleDynamics}


\subsubsection{Mass and momentum balances}


The horizontal momentum balance yields 
\beq
\label{eq:HorMomUpToOrderFive}
\nabla_\bfeta \pi\order{j}
  & = 
    & 0 \qquad (j = 0, ..., 5)
      \\
\label{u.1.sum} 
D_t\order{0} \bfu^{(1)} + f_{\perp}w^{(0)}\mathbf{e}_1 + \frac{1}{\Gamma}\nabla_\bfeta \pi^{(6)} 
  & = 
    & 0\,, 
\eeq
where $\pi$ is the Exner pressure defined in \eq{eq:ExnerDefinition}, and 
\beq
D_t\order{0}  = \pa_t  + \bfu^{(1)}\cdot \nabla_\bfeta + w^{(0)}\pa_z\,,
\qquad
\mathbf{e}_1 = (1,0)^T\,,
\eeq
and $\pi\order{6}$ satisfies the elliptic equation
\begin{align}
\frac{1}{\Gamma}\Delta_\bfeta \pi^{(6)}
= - \nabla_\bfeta\cdot 
        \left(\pa_t {\bfu}^{(1)}
            + \bfu^{(1)} \cdot\nabla_\bfeta{\bfu}^{(1)}
            + w^{(0)}\left(\pa_z \bfu^{(1)}+f_{\perp}\mathbf{e}_1\right)
        \right)\,.
\end{align}
An expression for 
$\nabla_\bfeta\cdot \pa_t {\bfu}^{(1)} = \pa_t (\nabla_\bfeta\cdot{\bfu}^{(1)})$
follows from the leading-order mass conservation equation, 
\beq
\pa_t (\nabla_\bfeta \cdot {\bfu}^{(1)})=( 1 -\pa_z)\pa_t w^{(0)}\,.
\eeq
where we use that pressure and density are dominated to leading order by 
$p\order{0} = \rho\order{0} = e^{-z}$ on account of \eq{eq:PHydroExpansion},
\eq{eq:PHydroExpansion}, and $\rho = p/T (1+\e^2(\qv+q_c+q_r))/(1+\e^{2}\qv/E)$, 
for the regime with $\alpha = 1$ from table~\ref{tab:DistinguishedLimit}.

Just as the pressure gradient alone dominates the horizontal momentum
balance up to $6$th order, see \eq{eq:HorMomUpToOrderFive}, the vertical
momentum balance is dominated by the pressure gradient up to fourth order 
and the accompanying gravity terms. In particular, at second order, the 
vertical pressure gradient and the gravity term are in balance. Since, 
furthermore, the pressure is horizontally homogeneous at that order, we 
subtract the balance in the environment of the cloud tower from the balance 
of the terms within to find zero total buoyancy at that order, 
\beq\label{theta.A.2}
\thetatilde^{(2)} = \theta\order{2} - \overline{\theta}^{(2)}
= -\Big(\Big(\frac{1}{E}-1\Big) 
\left(\qv^{(0)} - \overline{q}_v^{(0)}\right)  \ -\ \qc\order{0} - q_r^{(0)}\Big)\,.
\eeq
This balance will play a central role in what follows. 


\subsubsection{Saturated air}

Within the cloud tower, the air is by definition saturated with moisture, such that
the deviation of the leading order water vapor content satisfies 
\beq\label{eq:QvTildeZeroSat}
\qv\order{0} = q_{vs}^{(0)}(z)\,.
\eeq
The cloud water mixing ratio vanishes to leading order, \ie,
\beq
q_c=\e q_c^{(1)}+O(\e^2)
\eeq
due to rapid collection of cloud water by the falling rain, see the
term $-\e^{-1} S_{cr}$ in \eq{as.qc}. As a consequence,
the term $-\qc\order{0}$ in \eq{theta.A.2} vanishes identically. Also, 
the leading order source terms for cloud water must then balance in the 
equation for the cloud water mixing ratio, and this determines the first 
order cloud water content as a function of $w\order{0}$ and $\qr\order{0}$ 
through
\beq\label{qc1.sum}
 - w^{(0)}\frac{d q_{vs}^{(0)}}{dz} = S_{cr}^{(1)} = C_{cr}q_{c }^{(1)}q_{r }^{(0)}\,.
\eeq
Note that this relation also implies $w^{(0)}\geq 0$ in the saturated region 
since the mixing ratios must be positive: As a consequence, vertical upward 
motion is possible against the stable stratification as the latter is overcome 
by the release of latent heat. Yet, downward vertical motion is suppressed at 
leading order, because the cloud water content -- being rapidly washed out by 
precipitation --  is insufficient to overcome the stable stratification by 
re-evaporation in a downward motion. 

The total liquid water content is thus equivalent to the rain water content 
at leading order, and its mixture fraction obeys the transport equation
\beq
\label{eq:qr0sat}
D_t\order{0} {q}_{r}^{(0)}
-\frac{1}{\r^{(0)}} \pa_z(\r^{(0)}  q_{r }^{(0)}V_r)  
 = S_{cr}^{(1)} = -w^{(0)}\frac{d q_{vs}^{(0)}}{dz} \,.
\eeq

The buoyancy balance from \eq{theta.A.2}, with $\qc\order{0}$ eliminated and
with $\qv\order{0}, \overline{q}_v^{(0)}$ replaced with known functions of $z$, expresses 
the potential temperature perturbation as a function of $\qr\order{0}$ only. 
At the same time, however, $\thetatilde\order{2}$ must satisfy the second 
order potential temperature transport equation, 
\beq
D_t\order{0} \thetatilde\order{2}
= w\order{0} \frac{d\Delta\Theta^{(2)}}{dz} 
  + \qr\order{0} \left(V_r-w\order{0}\right) k_l\frac{dT^{(1)}}{dz}\,,
\eeq
where $\Delta\Theta^{(2)} = \Theta^{(2)}_{\rm ad} - \overline{\theta}^{(2)}$ is the difference between
the second order moist adiabatic and the second order background potential
temperature distributions. The determining equations for $\Theta^{(2)}_{\rm ad}(z)$ 
are worked out in appendix~\ref{sapp:MADFirstOrder}.

The rain water mixing ratio, $\qr\order{0}$, in turn satisifies \eq{eq:qr0sat} 
and combining these constraints yields an algebraic relation for the vertical 
velocity,%
\beq
w^{(0)} \biggl(k_l q_r^{(0)}\frac{dT^{(1)}}{dz} 
  & -
    & \frac{d\Delta\Theta^{(2)}}{dz}
      - \frac{1}{E}\frac{d q_{vs}^{(0)}}{dz}+\Big(\frac{1}{E}-1\Big)\frac{d\overline{q}_{v}^{(0)}}{dz}\biggr)
      \nonumber\\
\label{w0.sum}
  & = 
    & k_l q_r^{(0)} V_r \frac{d T^{(1)}}{dz} 
      - \frac{1}{\r_0}\pa_z \left( \r_0q_r\order{0}V_r \right)\,,
\eeq
where we recall that $T^{(1)}=\theta^{(1)}-\Gamma z$. 

We note that if $q_r^{(0)}\equiv 0$, then the vertical velocity vanishes as well. 
This means that on this long time scale under consideration, sustaining a vertical 
velocity is only possible if the system produces precipitation and, in turn, 
where there is no vertical velocity, no rain water can be found. 

\begin{figure}
\begin{center}

\includegraphics[width=0.49\textwidth]{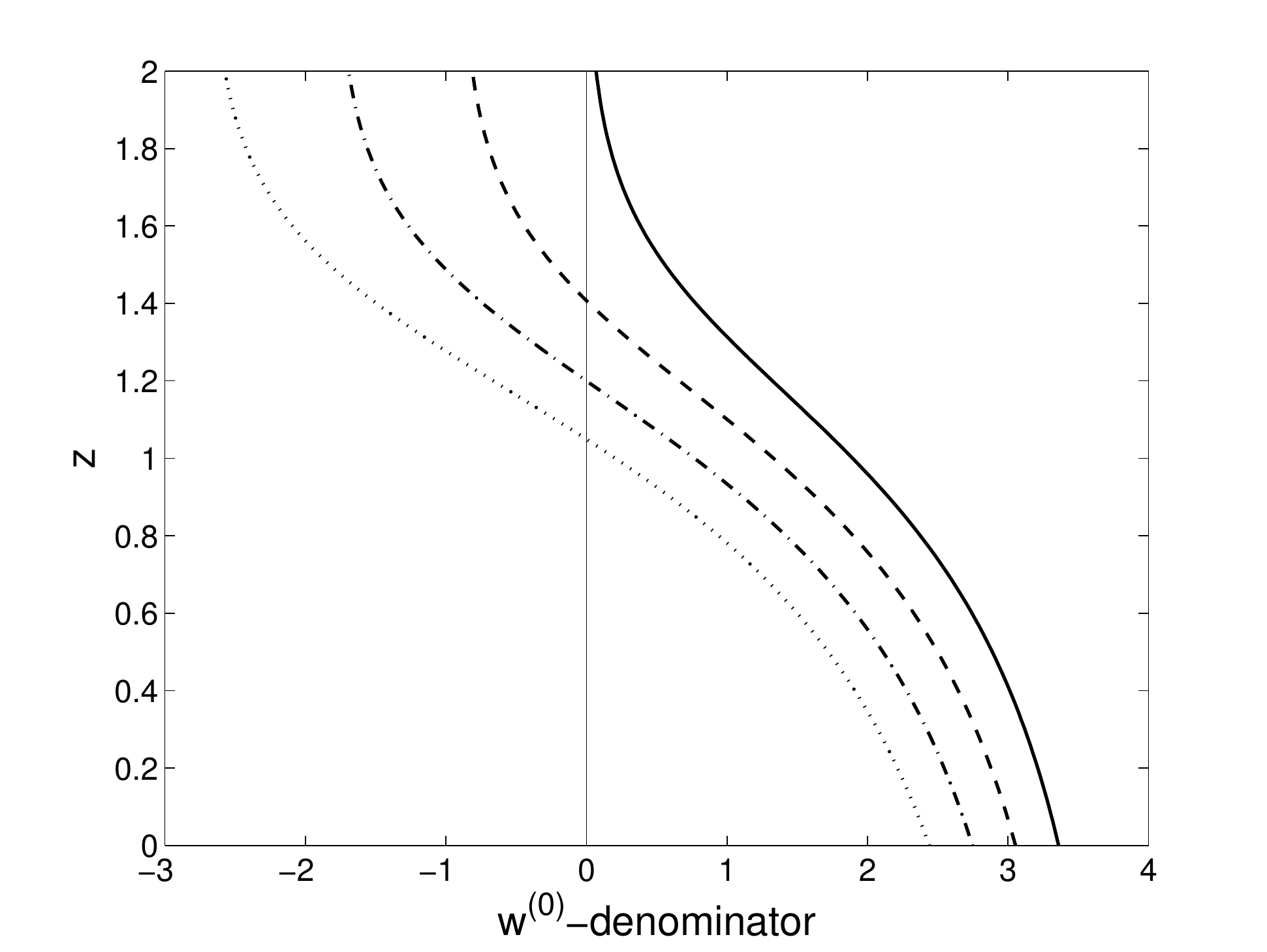}
\hfil\includegraphics[width=0.49\textwidth]{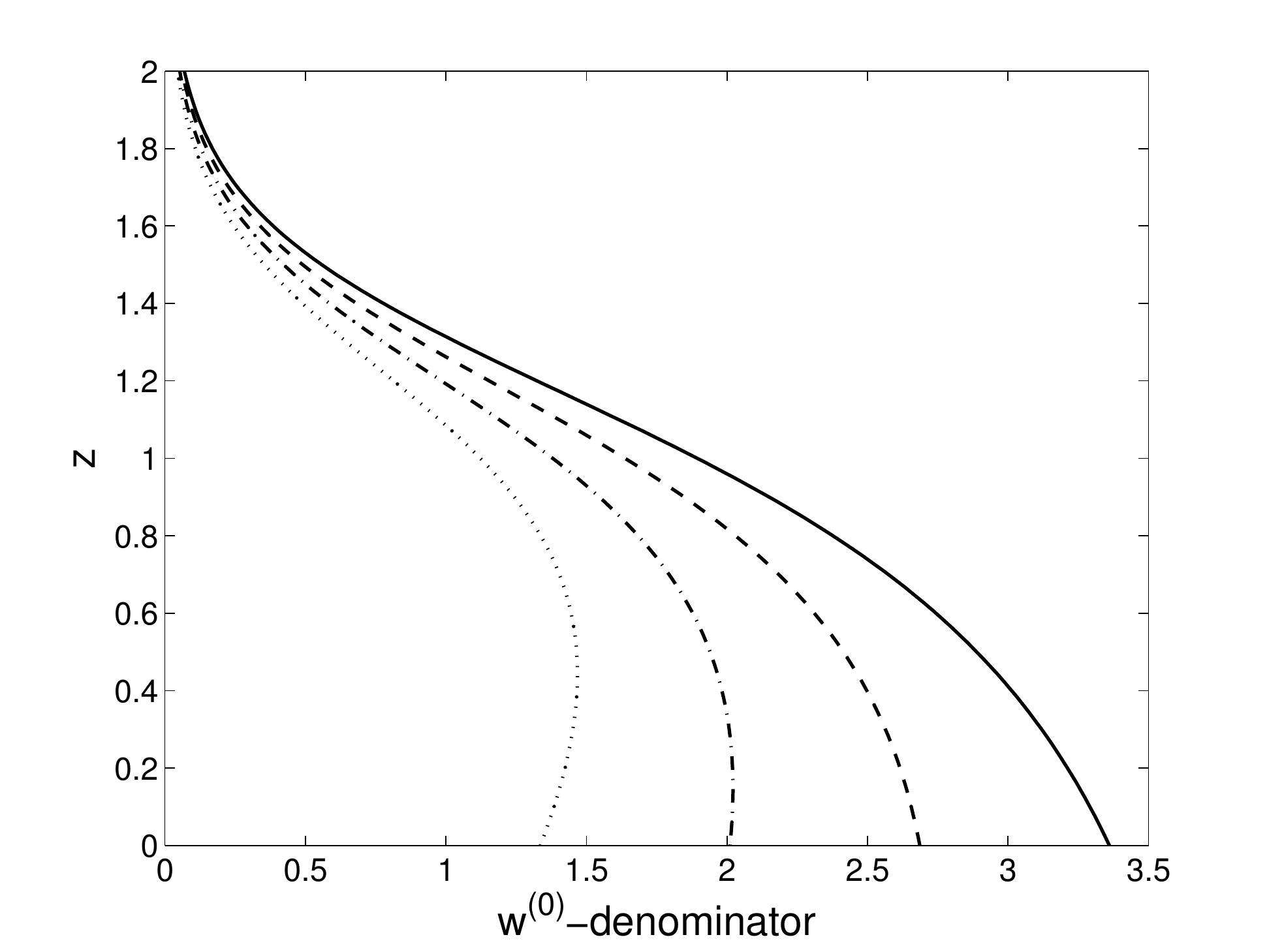}

\end{center}
\caption{The denominator in the determining equation \eq{fig:VerticalVelocitySaturated} 
for vertical velocity in saturated air for $\Delta\Theta_2 = 0$ and
$q_r^{(0)} = 2.2, 1.6, 0.8, 0.0$ (dotted, dash-dotted, dashed, and solid lines)} (left
panel), and for $q_r^{(0)} = (2.2, 1.6, 0.8, 0.0) \, q_{vs}^{(0)}(z)$ (right panel).
\label{fig:W0Denominator} 
\end{figure}

A rewrite of $w^{(0)}$ according to \eqref{w0.sum} reveals the following dependence 
on $q_r^{(0)}$ and $\pa_z q_r^{(0)}$
\beq\label{fig:VerticalVelocitySaturated}
w^{(0)}
= V_r 
  \frac{\left[k_l  \frac{d T^{(1)}}{dz} + 1\right]	 q_r^{(0)} - \pa_z q_r^{(0)}}%
  {k_l \frac{dT^{(1)}}{dz}q_r^{(0)} - \frac{d\Delta\Theta^{(2)}}{dz} 
    - \frac{1}{E}\frac{d q_{vs}^{(0)}}{dz}+\Big(\frac{1}{E}-1\Big)\frac{d\overline{q}_{v}^{(0)}}{dz}}\,.
\eeq
The denominator is rather benign for realistic values of $q_r^{(0)} \leq 2.4$ and
for dimensionless heights less than $1.0$, corresponding to a domain height of 
$10\, \textnormal{km}$, as shown in fig.~\ref{fig:W0Denominator}, left panel.
The right panel of  fig.~\ref{fig:W0Denominator} shows the denominator for the
more realistic setting where the rain water mixing ratio scales with the local
saturation water vapor mixing ratio. In this case, the denominator even stays 
positive throughout the bottom two scale heights of the atmosphere.

As a consequence, the qualitative properties of the $q_r$-equation will be dominated by  
the numerator. The explicit dependence on $q_r^{(0)}$ will induce a Burgers-type advective
nonlinearity, whereas the appearance of $\pa_z q_r^{(0)}$ induces a Hamilton-Jacobi-type term. 
In fact, with obvious abbreviations we have
\beq
w^{(0)} = a q_r^{(0)} - b \pa_z q_r^{(0)} 
\eeq
and, neglecting horizontal derivatives for simplicity, the equation for $q_r$ reads
\beq\label{eq:SaturatedHamiltonJacobi}
\pa_t q_r^{(0)} + \left(a q_r^{(0)} - b \frac{d q_{vs}^{(0)}}{dz} - V_r\right) \pa_z q^{(0)}_r - b (\pa_z q_r^{(0)})^2  = - \big(a\frac{d q_{vs}^{(0)}}{dz} +V_r \big) q_r^{(0)}\,.\quad
\eeq
%


\subsubsection{Undersaturated air}

In undersaturated regions within a narrow tower, all cloud water will rapidly 
evaporate, so that $\qc\order{0} \equiv 0$. Precipitation that descends into an 
undersaturated region will evaporate at a rate of order unity on the time scale
considered here, so that the remaining moisture variables $\qv, \qr$ to leading 
order satisfy the transport equations 
\beq\label{eq:DryTowerDynamicsI}
D_t\order{0}\qv\order{0}  
  & = 
    & S_{ev}\order{0}\,, 
      \\
D_t\order{0}\qr\order{0}-\frac{1}{\r_0}  \pa_z(\r_0 \qr\order{0}V_r)  
  & = 
    & - S_{ev}\order{0}\,,
\eeq
where
\beq\label{eq:DryTowerDynamicsII}
S_{ev}\order{0} = L  C_{ev} (q_{vs}^{(0)}-\qv\order{0})q_{r}\order{0}\,.
\eeq
In the undersaturated regions, the rather strong stability associated with the 
moist adiabatic potential temperature distribution is not overcome by matching
latent heat release from condensation as it is in the saturated region. As a
consequence, the vertical velocity is determined, as in the ``weak temperature
gradient approximation'' \cite{HeldHoskins1985,SobelEtAl2001,KleinMajda2006}, 
by the quasi-steady form of the potential temperature transport equation, 
\beq\label{eq:DryTowerDynamicsIII}
&&w^{(0)}\frac{d\theta^{(1)}}{dz}=-L S_{ev}^{(0)}=-L  C_{ev} (q_{vs}^{(0)}-q_v^{(0)})q_{r}^{(0)}\,.
\eeq 
%


\subsection{Differences between the moist thermodynamics scaling regimes}
\label{ssec:DifferencesBetweenRegimes}

Although the asymptotic approximations to the moist adiabatic distribution 
were comparably accurate for the two distinguished limit regimes from 
table~\ref{tab:DistinguishedLimit} (see fig.~\ref{fig:MoistAdiabat}), there are
subtle differences for the approximate dynamics of a narrow tower. Taking into 
account the different scaling regimes labelled $\alpha = 0$ and $\alpha = 1$ 
in table~\ref{tab:DistinguishedLimit}, we obtain for the expansion of the density 
potential temperature,
\beq\label{theta.rho}
\theta_\r &=& 1 + \e \Big(\theta^{(1)} + \frac{1-\alpha}{E} q_v^{(0)} \Big) \\
&&\qquad +\e^2 \Big(\theta^{(2)} + \frac{1-\alpha}{E} q_v^{(1)} + \big(\frac{\alpha}{E} -1\big)q_v^{(0)} - q_c^{(0)} -q_r^{(0)}\Big)\Big) +O(\e^3)\nonumber
\eeq
For $\alpha=0$ we obtain therefore from the vertical momentum balance to $O(\e^{-2})$ 
the additional condition 
\beq
0=-\frac{\widetilde \r^{(1)}}{p^{(0)}}=\widetilde\theta_\rho^{(1)}=\frac{1-\alpha}{E} \widetilde  q_v^{(0)}\,
\eeq
implying $q_v^{(0)}\equiv \bar q_v^{(0)} $. Since we want to allow for saturation 
at least within the core of a narrow cloud tower, this results in the condition
\beq\label{saturation}
q_v^{(0)}\equiv q_{vs}^{(0)}\,.
\eeq
This amounts to the air being close to saturation everywhere, which is 
common in the tropics. A distinction between saturated and undersaturated air is then 
made based upon the first order components, \ie,
\begin{quote}
\begin{center}
\begin{tabular}{lcl}
saturated
  & : 
    & $q_{v}^{(1)}\equiv q_{vs}^{(1)}$
      \\[5pt]
undersaturated
  & : 
    & $q_v^{(1)}<q_{vs}^{(1)}$ with $q_{vs}^{(1)}-q_v^{(1)}=O(1)$
\end{tabular}
\end{center}
\end{quote}
Since in the present hot tower setting the regime $\alpha = 0$ requires this 
restriction of almost saturation everywhere, we discuss here the differences
between both scaling regimes in this particular setting and assume \eqref{saturation} 
to hold throughout this subsection. The difference of the regimes then enters 
via the diagnostic relation from the buoyancy to second order

\beq\label{theta.rho.2}
0=\widetilde \theta_\r^{(2)} = \widetilde \theta^{(2)} + \frac{1-\alpha}{E} \widetilde q_v^{(1)}  - q_c^{(0)} -q_r^{(0)}\,.
\eeq


\subsubsection{Saturated air}
In saturated air, as mentioned above,  we have $q_{v}^{(1)}\equiv q_{vs,1}$. For the cloud water mixing ratio we obtain, in analogy with the earlier calculations
\beq
q_c^{(0)}=0\,,\qquad C_{cr} q_c^{(1)}q_r^{(0)} = -w^{(0)} \frac{d q_{vs}^{(0)}}{dz}\,.
\eeq
The leading order rain dynamics is the same for both scaling regimes as well,
\beq
D_t^{(0)}q_r^{(0)} -\frac{1}{\r_0}\pa_z (\r^{(0)} q_r^{(0)} V_r)= -w^{(0)} \frac{d q_{vs}^{(0)}}{dz}\,.
\eeq
Also the equation for the potential temperature fluctuation is again as before
\beq
D_t^{(0)} \widetilde \theta^{(2)} = w^{(0)} \frac{d\Delta \Theta^{(2)}}{dz} +  k_l q_r^{(0)} (V_r -w^{(0)}) \frac{dT^{(1)}}{dz}\,.
\eeq
Inserting now the balance equation from the buoyancy to second order  \eqref{theta.rho.2} we obtain different relations for the vertical velocity 
\beq
w^{(0)} \biggl(k_l q_r^{(0)}\frac{dT^{(1)}}{dz} 
  & - 
    & \frac{d\Delta\Theta^{(2)}}{dz}
      - \frac{1-\alpha}{E}\frac{d \Delta Q_{v}^{(1)}}{dz} - \frac{d q_{vs}^{(0)}}{dz}\biggr)
      \nonumber\\
\label{w0.sum.Alpha0}
  & = 
    & k_l q_r^{(0)} V_r \frac{d T^{(1)}}{dz} 
      - \frac{V_r}{\r_0}\pa_z \left( \r^{(0)} q_r^{(0)}\right)\,,
\eeq
where $ \Delta Q_{v}^{(1)}= q_{vs}^{(1)} - \overline{q}_{v}^{(1)}$. 


\subsubsection{Undersaturated air}

In undersaturated air we have $q_c\equiv 0$ and $q_v^{(1)}<q_{vs}^{(1)}$. Moreover we note that due to the condition of everywhere almost saturation in \eqref{saturation}, the evaporation vanishes to leading order and we have
\beq\label{evaporation}
S_{ev}^{(0)}=0 \,\qquad \textnormal{and} \qquad S_{ev}^{(1)} = C_{ev} (q_{vs}^{(1)}-q_v^{(1)})^+q_r^{(0)}\,.
\eeq
The strategy of obtaining the different solution components differs here from the previous setting for scaling regime 1. 
In particular we obtain the vanishing of the vertical velocity to leading order from  the equation for water vapor  using \eqref{saturation} and 
\eqref{evaporation} 
\beq
w^{(0)}\frac{dq_{vs}^{(0)}}{dz} =0 \qquad \textnormal{implying} \qquad w^{(0)}=0 \,.
\eeq
To next order we obtain
\beq\label{qv.1}
D_t^{(0)} q_v^{(1)} + w^{(1)} \frac{dq_{vs}^{(0)}}{dz} = S_{ev}^{(1)}\,.
\eeq
Due to the weak evaporation the rain water is also merely transported to leading order
\beq
D_t^{(0)} q_r^{(0)}-\frac{1}{\r_0}\pa_z (\r_0 q_r^{(0)} V_r)=0\,.
\eeq
To close the dynamics we still need to determine $w^{(1)}$, which we obtain again from the potential temperature equation. In the undersaturated region the latter reduces for $w^{(0)}=0$ to
\beq
D_t^{(0)} \widetilde\theta^{(2)} + w^{(1)} \frac{d \theta^{(1)}}{dz} = k_l q_r^{(0)} V_r\frac{d T^{(1)}}{dz} -  L S_{ev}^{(1)}\,.
\eeq
Note that averaging this equation in particular implies $\overline{w}^{(1)}=0$ and thus also $\pa_t\overline{q_{v}}^{(1)}=0$.
Therefore, using \eqref{theta.rho.2} and \eqref{eq:ThetaOneMADLeadingOrder} we can solve this equation for $w^{(1)}$ as follows
\begin{align}
\big(\frac{1-\alpha}{E}-L\big)\frac{dq_{vs}^{(0)}}{dz}\, w^{(1)}& &= \big( \frac{1-\alpha}{E}-L\big) S_{ev}^{(1)}  
+ V_r \Big(k_l \frac{dT^{(1)}}{dz}q_r^{(0)} -\frac{1}{\r_0}\pa_z (\r^{(0)} q_r^{(0)}) \Big)\,.
\end{align}


\section{Up- and downdrafts on the convection time scale}
\label{sec:UpAndDownDrafts}

Here we present sample numerical solutions for the up- and downdraft models 
derived in sections \ref{ssec:UpDrafts} and \ref{ssec:DownDrafts}, respectively. We 
restrict to the simplest settings, neglecting horizontal advection within the towers 
as well as (turbulent) transport, to reveal the essential behavior of the convective 
scale dynamics equations. The construction and investigation of a self-consistent 
tower model in which both regimes will be coupled by turbulent transport, and a 
thorough comparison with existing turbulent plume and buoyant bubble models for 
individual deep convection events is left for future work. 


\subsection{Updrafts}
\label{ssec:UpDrafts}

Here we solve the Hamilton-Jacobi type equation \eq{eq:SaturatedHamiltonJacobi} for
updrafts in saturated parts of a tower and for the asymptotic scaling regime 
$\alpha = 1$ from table~\ref{tab:DistinguishedLimit}, 
\beq\label{eq:SaturatedHamiltonJacobiRepeat}
\pa_t q_r^{(0)} + \left(a q_r^{(0)} - b \frac{q_{vs}^{(0)}}{dz} - V_r\right) \pa_z q^{(0)}_r - b (\pa_z q_r^{(0)})^2  = - \big(a\frac{q_{vs}^{(0)}}{dz} +V_r \big) q_r^{(0)}\,,\quad
\eeq
where the coefficients $a,b$ are given by
\beq
a 
  & = 
    &  \frac{V_r}{D} \left[k_l \frac{dT\order{1}}{dz} + 1\right]\,,
\qquad
b = \frac{V_r}{D} 
      \\
\noalign{\textnormal{with}}
D
  & = 
    & k_l \frac{dT^{(1)}}{dz}q_r^{(0)} - \frac{d\Delta\Theta^{(2)}}{dz} 
    - \frac{1}{E}\frac{d q_{vs}^{(0)}}{dz}+\Big(\frac{1}{E}-1\Big)\frac{d\overline{q}_{v}^{(0)}}{dz} \,.
\eeq
The equation is solved using Strang splitting between the Hamilton-Jacobi terms
involving the vertical derivative $\pa_z \qr\order{0}$ on the left, and the source
term proportional to $\qr\order{0}$ on the right. For the first split step we have 
adapted the first-order finite difference scheme for Hamilton-Jacobi equations by 
Crandall-Lions, \cite{CrandallLions1984}. The second split step has an obvious
analytical solution, namely 
\beq
{\qr}^{n+1}_i 
  = {\qr}^{n}_i\exp\left(-\left[a\, \frac{d\qvs}{dz} + V_r\right]\Delta t\right)\,,
\eeq
where we have dropped the $\order{0}$ superscript for convenience of notation, and
where ${\qr}^{n}_i$ denotes the approximate numerical value for $\qr\order{0}$
at time level $t^n = n \Delta t$ and grid location $z_i = i \Delta z$.

\begin{figure}[htbp]
\begin{center}
\begin{minipage}{0.45\textwidth}
\begin{center}
$w = W(\pa_z\qr, \qr, z); \quad 0 \leq t \leq 4.2$  

\medskip

\includegraphics[width=\textwidth]{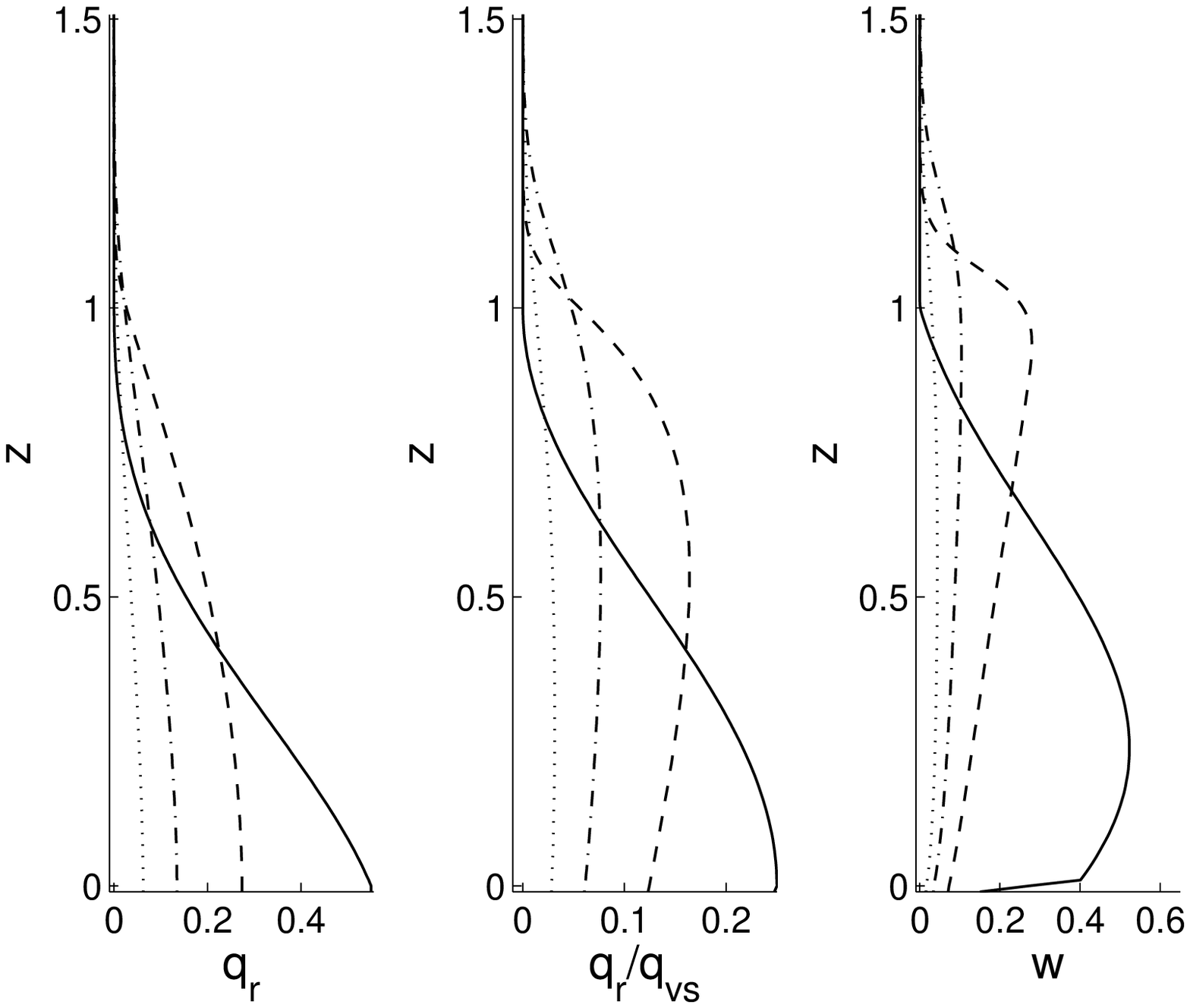}
\end{center}
\end{minipage}
\hfill
\begin{minipage}{0.45\textwidth}
\begin{center}
$w \equiv 0; \quad 0 \leq t \leq 0.8$ 

\medskip

\includegraphics[width=\textwidth]{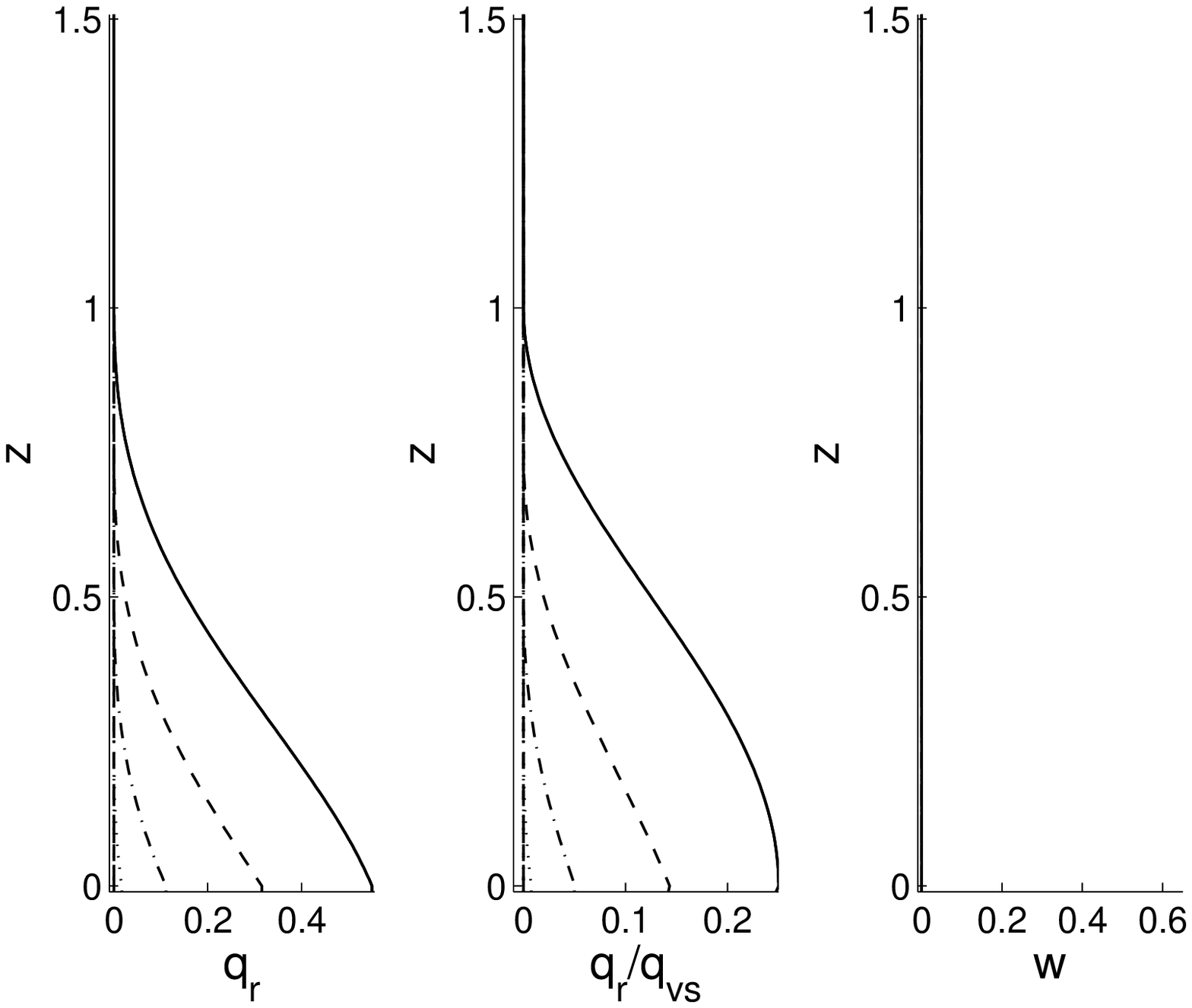}
\end{center}
\end{minipage}

\caption{Evolution of rain water mixing ratio with (left column) and without (right column)
the self-induced vertical velocity from \eq{fig:VerticalVelocitySaturated}. Output times 
in the left set of graphs: solid line $t = 0$, dashed $t = 1.378$, dash-dotted $t = 2.75$, 
dotted $t = 4.13$. Output times on the right: $t = 0$, dashed $t = 0.26$, 
dash-dotted $t = 0.51$, dotted $t = 0.77$.}
\label{fig:Updraft}
\end{center}
\end{figure}

Figure~\ref{fig:Updraft} gives an impression of the implications of the saturated 
tower dynamical equations by comparing the rain water dynamics with and without 
the self-induced vertical velocity from \eq{fig:VerticalVelocitySaturated}.
Both simulations start from initial data
\beq
\qr(0,z) 
= 
\left\{
\begin{array}{l@{\quad}l}
0.125 \, \qvszero(z)
\left(1 + \cos(\pi z)\right)
  &  (0 \leq z \leq 1)
    \\[10pt]
0
  & \textnormal{otherwise}
\end{array}
\right.
\eeq

The left triple of graphs
shows snapshots of vertical profiles of $\qr\order{0}$, $\qr\order{0}/\qvszero$, 
and $w\order{0}$ as they evolve under eq.~\eq{eq:SaturatedHamiltonJacobiRepeat} 
for times $0 \leq t \leq 4.2$. The right triple of graphs shows similar snapshots 
when the selfinduced vertical velocity is set to zero, so that the rain water 
simply precipitates with the terminal sedimentation velocity $-V_r$. The rain falls 
down rather rapidly in this case, so that we show snapshots within the interval 
$0 \leq t \leq 0.8$ in these graphs.  
Comparing the left and right sets of graphs we observe that the self-induced updraft 
tends to substantially prolong the life time of a convective tower, and this also 
implies much higher precipitation yield. 


\subsection{Downdrafts}
\label{ssec:DownDrafts}

Here we provide an example of the evolution of the water constituents and the 
vertical velocity in undersaturated regions of a cloud tower following 
eqs.~\eq{eq:DryTowerDynamicsI}--\eq{eq:DryTowerDynamicsIII}. The calculations
start from initial data
\beq
\left.
\begin{array}{rcl}
\qr(0,z) 
  & =
    & 0.25 \, \qvszero(z) (1-\cos(4 \pi z  / 3))
      \\[10pt]
\qv(0,z) 
  & =
    & 0.25\, \qvszero(z) 
\end{array}
\right\}
\qquad \left(0 \leq z \leq 3/2\right)
\eeq
and cover a rather short time interval of $0 \leq t \leq 0.4$.
\begin{figure}[htbp]
\begin{center}

\includegraphics[width=0.5\textwidth]{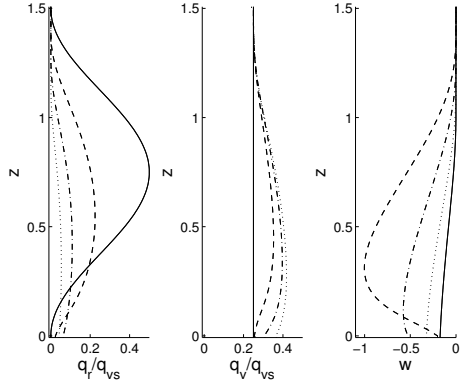}

\caption{Evolution of rain water mixing ratio, ... in undersaturated air. 
Output times: $t = 0$, dashed $t = 0.13$, 
dash-dotted $t = 0.25$, dotted $t = 0.38$.}
\label{fig:Downdraft}
\end{center}
\end{figure}
As expected, we see the precipitation descend and evaporate in the rightmost graph.
It thereby moistens the atmosphere as seen in the middle graph where the relative
humidity $\qv/\qvs$ increases by a factor of roughly two in the lower parts of the
domain in the course of time. The descend of the rain water is pronounced here
further in comparison with the saturated, updraft-free case discussed in the context
of fig.~\ref{fig:Updraft}, right triple of graphs, by the downdraft velocity induced 
by evaporative cooling (right-most graph in fig.~\ref{fig:Downdraft}). This is why
the present process is already completed essentially after dimensionless times of 
order $t \sim 0.5$.


\section{Conclusions}
\label{sec:Conclusions}

In this paper we have presented two alternative scaling regimes that allow 
us to incorporate a familiar class of bulk moist microphysics closures in
the general multiscale asymptotic modelling framework for atmospheric
flows summarized in \cite{Klein2010}. A first application of these, quite 
similar, scaling regimes to the dynamics of convective hot towers revealed a
mechanism of self-sustainance of precipitating updrafts in saturated air, and 
it provided an asymptotic description of strong downdrafts due to evaporative 
cooling in undersaturated air. 

This first application is as yet rudimentary, however, since we have set 
aside the issue of stability of the quasi-steady balances that characterize 
the considered flow regimes, three-dimensional advection within the
cloud towers, turbulent transport, lateral entrainment, the interaction 
of adjacent cloud towers, the influence of bottom boundary layers, and 
the large-scale organization of ensembles of hot towers. All these 
aspects shall be addressed in forthcoming publications.

\begin{acknowledgements}
The authors thank Olivier Pauluis (Courant Institute) for helpful discussions on the 
thermodynamics of moist air, and the Institute for Pure and Applied Mathematics (IPAM) 
at UCLA for hosting the Long Term Program 
\emph{Model and Data Hierarchies for Simulating and Understanding Climate} in 2010. 
In the course of this program the authors were able to lay the foundations for the 
present collaboration. S.H.\ thanks the Austrian Science Fund for their support via 
the Hertha-Firnberg project T-764. R.K.\ acknowledges support by the Deutsche 
Forschungsgemeinschaft through the Collaborative Research Center CRC 1114 
``Scaling Cascades in Complex Systems'', Project C06. The authors gratefully 
acknowledge the fabulous work of developers and maintainers of the free LaTeX word 
processing system and of the TeXShop TeX-writing environment. 
\end{acknowledgements}


\appendix


\section{Derivation of the scaled moist adiabatic equation \eq{eq:DThetaDzFinal}}
\label{app:ScaledMoistAdiabaticEquation}


\subsection{Effective equation for the moist adiabat}
\label{sapp:EffectiveMADEquation}

Inserting the hydrostatic equation \eq{eq:DPiDzMoistAdiabaticGen} into 
\eq{eq:DThetaDzMoistAdiabaticGen}, the equations for the moist adiabat
become 
\beq
\label{eq:DThetaDzMoistAdiabaticGenApp}
\frac{d\ln\theta}{dz}
  & =
    & - \e^{e-a} \frac{L\phi_\e(T)}{C_\e\pi\theta} \frac{d}{dz}\qvs(\theta,\pi;\e)
      \ + \ \e^{e}\frac{R_\e \Psi_\e(\qvs)}{C_\e} \frac{1}{\pi\theta} 
      \\
\label{eq:DPiDzMoistAdiabaticGenApp}
\frac{d\ln\pi}{dz}
  & =
    & - \e^b \frac{\Gamma}{\pi\theta} \Psi_\e(\qvs)
\eeq
where
\beq
\Psi_\e(\qvshat)
  & = 
    & \frac{1 + \e^e \qvs}{1 + \e^{e-c} \qvs/E}
      \\
C_\e
  & =
    & 1 + \e^{e+b-b_v-c}\frac{\Gamma}{E\Gamma_v}\qvs
      \\
R_\e
  & =
    & \e^{b-c} \frac{\Gamma}{E}\Big(\e^{b-b_v}\frac{\Gamma}{\Gamma_v} -1\Big) \qvs
\eeq
In the sequel we are interested in the dominant contributions in these equations, 
so we keep only the leading order terms. That is, we drop the second term on the right 
in \eq{eq:DThetaDzMoistAdiabaticGenApp}, and set $C_\e = \Psi_\e \equiv 1$ in the first 
terms of \eq{eq:DThetaDzMoistAdiabaticGenApp} and \eq{eq:DPiDzMoistAdiabaticGenApp} and 
obtain
\beq
\label{eq:DThetaDzMoistAdiabaticTrunc}
\frac{d\ln\theta}{dz}
  & =
    & - \e^{e-a} \frac{L\phi_\e(T)}{\pi\theta} \frac{d}{dz}\qvshat(\theta,\pi;\e)
      \\
\label{eq:DPiDzMoistAdiabaticTrunc}
\frac{d\ln\pi}{dz}
  & =
    & - \e^b \frac{\Gamma}{\pi\theta} 
\eeq
This is the starting point for the subsequent scale analysis. 

For the saturation water vapor mixing ratio, $q_{vs}$, the dependence on 
the unknowns $\theta, \pi$ follows from the definition of $\es$ in 
\eq{eq:QvsEsScaling} which depends on temperature only, so that -- 
with $T = \pi\theta$ --
\beq
\qvshat(\theta,\pi;\e) 
= \frac{E \eshat(T;\e)}{p - \e^{e-c}\eshat(T;\e)}
\equiv \frac{E \eshat(\theta\pi;\e)}{\pi^{1/\e^b\Gamma} - \e^{e-c}\eshat(\theta\pi;\e)}
\eeq
where the Clausius-Clapeyron relation for $\eshat$ reads as 
\beq\label{eq:ClausiusClapeyronScaled2App}
\frac{1}{\eshat}\frac{d\eshat}{dT} = \frac{1}{\e^{d}} \frac{AL}{T^2}\phi_\e(T)
\eeq
with $\phi_\e(T)$ from \eq{eq:LatentHeatTempDependence}. Evaluation of 
$d\qvshat/dz$ on the r.h.s.\ of \eq{eq:DThetaDzMoistAdiabaticGenApp} yields
\beq
\frac{d\qvshat}{dz} 
  & = 
    & \left[\frac{\pa \qvshat}{\pa\pi} \frac{d\pi}{dz}
          + \frac{\pa \qvshat}{\pa\theta} \frac{d\theta}{dz}
      \right]\,.
\eeq
Re-inserted into \eq{eq:DThetaDzMoistAdiabaticGenApp} and using 
\eq{eq:DPiDzMoistAdiabaticGenApp} we find
\beq\label{eq:DThetaDzMoistAdiabatic2Gen}
\left[1 + \e^{e-a}\frac{L\phi_\e}{\pi}\frac{\pa \qvshat}{\pa\theta}\right]\frac{d\theta}{dz}
= \e^{e-a}\frac{L\phi_\e}{\pi} \frac{\pa \qvshat}{\pa\pi} \frac{\e^b\Gamma}{\theta}\,.
\eeq
Using \eq{eq:ClausiusClapeyronScaled2}, the two partial derivatives of $q_{vs}$ become
\beq
\frac{\pa \qvshat}{\pa\theta} 
  & = 
    & \left[\frac{\qvshat}{\eshat} + \frac{\e^{e-c}\qvshat}{p-\e^{e-c}\eshat}\right] 
      \frac{\pa \eshat}{\pa\theta}
      = \left[\frac{\qvshat}{\eshat} + \frac{\e^{e-c}\qvshat}{p-\e^{e-c}\eshat}\right] 
        \pi \frac{d \eshat}{d T}  
    \nonumber\\ 
\label{eq:DQvsDThetaGen}
  & = 
    & \left[\frac{\qvshat}{\eshat} + \frac{\e^{e-c}\qvshat}{p-\e^{e-c}\eshat}\right] 
      \frac{\pi \eshat}{\e^{d}}\frac{AL\phi_\e}{T^2}  
      = \left[\frac{p}{p-\e^{e-c}{\widehat e}_s}\right]
        \frac{\pi}{\e^{d}}\frac{AL\phi_\e}{T^2} \qvshat
      \\
  & =
    & \bigoh{\e^{-d}} \qquad(\e \to 0)
      \nonumber\\[10pt]
\frac{\pa \qvshat}{\pa\pi} 
  & = 
    & \left[-\frac{\qvshat}{p-\e^{e-c}\eshat}\right] \frac{d p}{d\pi} 
      + \left[\frac{\qvshat}{\eshat} + \frac{\e^{e-c}\qvshat}{p-\e^{e-c}\eshat}\right] 
      \frac{\pa \eshat}{\pa\pi}
    \nonumber\\ 
  & = 
    & \left[- \frac{\qvshat}{p-\e^{e-c}\eshat}\right] \frac{d \pi^{1/\e^b\Gamma}}{d\pi} 
      + \left[\frac{\qvshat}{\eshat} + \frac{\e^{e-c}\qvshat}{p-\e^{e-c}\eshat}\right] \theta \frac{d \eshat}{dT}
    \nonumber\\ 
\label{eq:DQvsDPiGen}
  & = 
    & \left[\frac{p}{p-\e^{e-c}\eshat}\right] 
      \left(\frac{\theta}{\e^{d}}\frac{AL\phi_\e}{T^2}
           -\frac{1}{\e^b\Gamma\pi}
      \right) \qvshat
    \\ 
  & =
    & \bigoh{\e^{-b}, \e^{-d}} \qquad(\e \to 0)
\nonumber
\eeq
Collecting the results from \eq{eq:DThetaDzMoistAdiabatic2Gen}, 
\eq{eq:DQvsDThetaGen}, and \eq{eq:DQvsDPiGen} we have
\beq
\frac{d\theta}{dz} 
  & = 
    & \e^b \frac{\Gamma}{\theta}
      \frac{\left[\frac{p}{p-\e^{e-c}{\widehat e}_s}
            \right] \left(\frac{\theta}{\e^{d-e}} \frac{AL\phi_\e}{T^2}
          -  \frac{1}{\e^{b-e}}\frac{1}{\Gamma\pi}\right) \, {\widehat q}_{vs}}%
      {\left[\frac{p}{p-\e^{e-c}{\widehat e}_s}
          \right] \frac{\pi}{\e^{d-e}}\frac{AL\phi_\e}{T^2} \qvshat + \frac{\e^a\pi}{L\phi_\e}}
      \nonumber\\
  & = 
    & \e^b \frac{\Gamma}{\pi}
      \frac{\left(\frac{AL\phi_\e}{T^2}
          - \frac{\e^{d-b}}{\Gamma\theta\pi}\right) \qvshat}%
      {\frac{AL\phi_\e}{T^2}\qvshat 
       + \frac{\e^{a+d-e} I_\e }{L\phi_\e}} 
          \qquad 
          \left(I_\e \equiv 1 - \frac{\e^{e-c}\eshat}{p}\right)
      \nonumber\\
\label{eq:DThetaDzFinalApp}
  & =  
    & \e^b \frac{\Gamma}{\pi}
      \left( 1 - \e^{a+d-e}
        \frac{I_\e
            + \e^{e-a-b} \frac{L\phi_\e}{\Gamma T} \qvshat}%
        {\frac{AL^2\phi^2_\e}{T^2} \qvshat + \e^{a+d-e} I_\e}
      \right)
\eeq

The hindsight check of validity of the truncations we introduced in going from
\eq{eq:DThetaDzMoistAdiabaticGenApp}, \eq{eq:DPiDzMoistAdiabaticGenApp} to the 
simpler system \eq{eq:DThetaDzMoistAdiabaticTrunc}, \eq{eq:DPiDzMoistAdiabaticTrunc}, 
given the final results for the various exponents in 
\eq{eq:ThermodynamicScalingsGen} of the scale analysis in 
section~\ref{sec:MoistAdiabaticScalings}, \ie, 
\beq
a = b = b_v = c = 1\,, \qquad\textnormal{and}\qquad e = 2
\eeq
shows that $\e^e \ll \e^{e-a}$, $\e^{e+b-b_v-c} = \e^{e-c} = \e \ll 1$, and that
justifies the approximations introduced. 


\section{Asymptotics of the moist adiabat}
\label{app:MADAsymptotics}


\subsection{Expansion of the balance equations for the moist adiabat}
\label{sapp:MADExpansion}

We recall that scaling regime 1 amounts to $\alpha=1$ and scaling regime 2 to 
$\alpha=0$. The following derivations are valid for both regimes, where in the 
expansions we account for the different regimes by making use of the switching 
function $(1-\alpha)$ as a prefactor for terms, which vanish in regime 1, but are 
present in regime 2. For the moist adiabat (i.e. at saturation without liquid water) 
we have the balance equations 
\beq
\label{eq:DThetaDzMoistAdiabaticGenAppAsymp}
\frac{d\theta}{dz}
  & =
    & - \e \frac{L\phi_\e(T)}{C_\e\pi} \frac{d}{dz}\qvs(\theta,\pi;\e)
      + \e^{2}\frac{R_\e \Psi_\e(\qvs)}{C_\e \pi}  
      \\
\label{eq:DPiDzMoistAdiabaticGenAppAsymp}
\frac{d\pi}{dz}
  & =
    & - \e \frac{\Gamma}{\theta} \Psi_\e(\qvs)
\eeq
where
\beq
\Psi_\e(\qvs) 
  & = 
    & \frac{1 + \e^2 \qvs}{1 + \e^{1+\alpha}\, \qvs/E}
      \\
C_\e
  & =
    & 1 + \e^{1+\alpha}\, k_v \qvs
      \\
R_\e
  & =
    & \e^\alpha \kappa_v \qvs
\eeq
We insert the expansions 
\beq
\theta 
  & = 
    & 1 + \e\theta\order{1} + \e^2 \theta\order{2} + \littleoh{\e^2}
      \\
\pi 
  & = 
    & 1 + \e\pi\order{1} + \e^2 \pi\order{2} + \littleoh{\e^2}
      \\
\qvs 
  & = 
    & q_{vs}^{(0)} + \e q_{vs}^{(1)} + \littleoh{\e}\,,
\eeq
of the unknowns into \eq{eq:DThetaDzMoistAdiabaticGenAppAsymp}, 
\eq{eq:DPiDzMoistAdiabaticGenAppAsymp} to obtain
\beq
\label{eq:DThetaOneDzMoistAdiabaticApp}
\frac{d\theta\order{1}}{dz}
   & =
     & - L \frac{d q_{vs}^{(0)}}{dz}
       \\
\label{eq:DPiOneDzMoistAdiabaticApp}
\frac{d\pi\order{1}}{dz} 
  & =
    & - \Gamma 
      \\
\label{eq:DThetaTwoDzMoistAdiabaticApp}
\frac{d\theta\order{2}}{dz}
  & =
    & - L \frac{d q_{vs}^{(1)} }{dz} 
      + L \frac{d q_{vs}^{(0)}}{dz} 
        \left(\chi T\order{1} +\pi^{(1)}+ (1-\alpha) k_v q_{vs}^{(0)}\right)
      + (1-\alpha) \kappa_v q_{vs}^{(0)}\, .
      \\
\label{eq:DPiTwoDzMoistAdiabaticApp}
\frac{d\pi\order{2}}{dz} 
  & =
    & \Gamma \left(\theta\order{1} + (1-\alpha) q_{vs}^{(0)}/E\right)
\eeq
To close \eq{eq:DThetaOneDzMoistAdiabaticApp} we recall the constitutive equation 
for $\qvs$, \ie, 
\beq\label{eq:QvsEsRelation}
\qvs = \frac{E \es(T)}{p - \e^{1+\alpha} \es(T)}
\qquad\textnormal{where $\es(T)$ satisfies}\qquad
\frac{d\ln\es}{dT} =  \frac{AL}{\e} \frac{\phi_\e(T)}{T^2}\,.
\eeq
This leads to
\beq
\frac{d\qvs}{dz} 
  & = 
    & \left[\frac{\qvs}{\es} + \frac{\e^{1+\alpha}\qvs}{p - \e^{1+\alpha}\es}\right] 
      \frac{d\es}{dT}\frac{dT}{dz} 
      - \left[\frac{\qvs}{p - \e^{1+\alpha}\es}\right] \frac{dp}{dz}
      \nonumber\\
  & = 
    & \qvs \left[1 + \frac{\e^{1+\alpha}\es}{p - \e^{1+\alpha}\es}\right] \frac{d\ln\es}{dT}\frac{dT}{dz} 
      - \qvs \left[\frac{p}{p - \e^{1+\alpha}\es}\right] \frac{d\ln p}{dz}
      \nonumber\\
  & = 
    & \qvs \left[\frac{1}{1 - \e^{1+\alpha}\es/p}\right] \left( 
       \frac{AL}{\e}\frac{\phi_\e(T)}{T}\frac{d\ln T}{dz} 
      - \frac{1}{\e\Gamma}\frac{d\ln \pi}{dz}
      \right)
      \\
  & = 
    & \qvs \left[\frac{1}{1 - \e^{1+\alpha}\es/p}\right] \left( 
       \frac{AL\phi_\e(T)}{T}\left(\frac{1}{\theta}\frac{d\thetatilde}{dz} 
                                 +  \frac{1}{\pi}\frac{d\pitilde}{dz}\right)
      - \frac{1}{\Gamma\pi}\frac{d\pitilde}{dz}
      \right)
      \nonumber
\eeq
where we have used that $T = \pi\theta$ and introduced the abbreviations 
$\theta = 1 + \e\thetatilde$ and $\pi = 1 + \e\pitilde$. Expanding this result 
to leading and first order we find
\beq
\label{eq:DQvsNullDzApp}
\frac{dq_{vs}^{(0)}}{dz}
  & =
    & q_{vs}^{(0)} 
      \left( AL \frac{d\theta\order{1}}{dz} 
            + \frac{AL\Gamma - 1}{\Gamma}\frac{d\pi\order{1}}{dz} \right)
      \\
\label{eq:DQvsOneDzApp1}
\frac{d\qvs\order{1}}{dz}
  & =
    & \left(\qvs\order{1} + (1-\alpha) q_{vs}^{(0)} \left(\frac{\es}{p}\right)\order{0} \right) 
      \left( AL \frac{d\theta\order{1}}{dz} + \frac{AL\Gamma - 1}{\Gamma}\frac{d\pi\order{1}}{dz} 
      \right)
      \nonumber\\
  & 
    & + \ q_{vs}^{(0)} 
      \left( AL \frac{d\theta\order{2}}{dz} 
        + \frac{AL\Gamma - 1}{\Gamma}\frac{d\pi\order{2}}{dz}
      \right) 
      \\
  & 
    & - \ q_{vs}^{(0)}
      AL \, (\chi+1) \,T\order{1} \left(\frac{d\theta\order{1}}{dz} + \frac{d\pi\order{1}}{dz}\right)
      \nonumber\\
  & 
    & + \ q_{vs}^{(0)} 
       \left( - AL \left[\theta\order{1}\frac{d\theta\order{1}}{dz} 
                         + \pi\order{1}\frac{d\pi\order{1}}{dz} 
                   \right] +  \frac{\pi\order{1}}{\Gamma} \frac{d\pi\order{1}}{dz}\right)
      \nonumber
\eeq
%


\subsection{Leading order moist adiabatic solution}
\label{sapp:MADLeadingOrder}

Equation \eq{eq:DPiOneDzMoistAdiabaticApp} is readily solved by 
\beq
\pi\order{1} = -\Gamma z\,.
\eeq
Re-inserting \eq{eq:DThetaOneDzMoistAdiabaticApp} into the 
\eq{eq:DQvsNullDzApp} we find a closed equation for $q_{vs}^{(0)}$, 
\beq
\label{eq:QvsNullEquation}
\frac{dq_{vs}^{(0)}}{dz}
  & =
    & - \frac{(AL\Gamma - 1) q_{vs}^{(0)}}{AL^2 q_{vs}^{(0)} + 1}
\eeq
which is readily solved by \eq{eq:QvsZeroMADLeadingOrder}, \ie, 
\beq
\label{eq:QvsNullSolution}
\ln \left(\frac{q_{vs}^{(0)}}{{q_{vs}^{(0)}}_0}\right) 
+ AL^2\, \left(q_{vs}^{(0)} - {q_{vs}^{(0)}}_0\right) 
= - (AL\Gamma - 1) z \, ,
\eeq
With these results we also have
\beq
\label{eq:ThetaOneOfQvs}
\theta\order{1} 
  & =
    & -L \left(q_{vs}^{(0)}(z) - q_{vs}^{(0)}(0)\right)
      \\ 
T\order{1}
  & = 
    & \theta\order{1} + \pi\order{1} = \theta\order{1} - \Gamma z\,.
\eeq
%


\subsection{First order moist adiabatic solution}
\label{sapp:MADFirstOrder}

Next we reconsider \eq{eq:DQvsOneDzApp1} using \eq{eq:DQvsNullDzApp} and 
\eq{eq:QvsNullEquation} to obtain
\beq
\label{eq:DQvsOneDzApp2}
\frac{d\qvs\order{1}}{dz}
  & =
    & \left(\frac{\qvs\order{1}}{q_{vs}^{(0)}} 
         + ( 1-\alpha )\left(\frac{\es}{p}\right)\order{0} \right) 
      \frac{dq_{vs}^{(0)}}{dz}
      \nonumber\\
  & 
    & + \ q_{vs}^{(0)} 
      \left( AL \frac{d\theta\order{2}}{dz} 
        + \frac{AL\Gamma - 1}{\Gamma}\frac{d\pi\order{2}}{dz}
      \right) 
      \\
  & 
    & - \ q_{vs}^{(0)}\left(
      \frac{AL}{2} 
      \left[ (\chi+1)  \frac{d{T\order{1}}^2}{dz} + \frac{d{\theta\order{1}}^2}{dz}\right]
      + \frac{AL\Gamma - 1}{2\Gamma}\frac{d{\pi\order{1}}^2}{dz}\right)
      \nonumber
\eeq
This is recast collecting the terms involving $\qvs\order{1}$, $\theta\order{2}$, and
$\pi\order{2}$ on the left and using that \eq{eq:QvsEsRelation} can be solved for $\es/p$ 
to yield
\beq
\frac{\es}{p} = \frac{\qvs}{E + \e^{1+\alpha} \qvs}
\qquad\textnormal{\ie}\qquad
\left(\frac{\es}{p}\right)\order{0} = \frac{q_{vs}^{(0)}}{E}\,.
\eeq
After division of \eq{eq:DQvsOneDzApp2} by $q_{vs}^{(0)}$ and reordering terms, we find
\beq
\frac{d}{dz}\left(\frac{\qvs\order{1}}{q_{vs}^{(0)}}\right)
  & - 
    & AL \frac{d\theta\order{2}}{dz} - \frac{AL\Gamma - 1}{\Gamma}  \frac{d\pi\order{2}}{dz}
\label{eq:DQvsOneDzApp}
      \\
  & =
    & (1-\alpha) \frac{1}{E}\frac{dq_{vs}^{(0)}}{dz} 
      - \ \frac{d}{dz}
          \left(
            \frac{AL}{2} 
            \left[ (\chi+1)  {T\order{1}}^2 + {\theta\order{1}}^2\right]
          + \frac{AL\Gamma - 1}{2\Gamma} {\pi\order{1}}^2\right)
      \nonumber
\eeq
Integrating \eq{eq:DQvsOneDzApp} w.r.t.\ $z$ we obtain a first expression involving
$\qvs\order{1}$, $\theta\order{2}$, and $\pi\order{2}$
\beq
\label{eq:ThetaTwoQvsOneEqn1}
\hspace{-10pt}\frac{\qvs\order{1}}{q_{vs}^{(0)}} - AL \theta\order{2}
- \frac{AL\Gamma - 1}{\Gamma} \pi\order{2} 
  & =
    & (1-\alpha) \frac{1}{E} \left({q_{vs}^{(0)}} - {q_{vs}^{(0)}}_0\right)
      \nonumber\\
  & 
    & - \frac{AL\Gamma - 1}{2\Gamma} {\pi\order{1}}^2 
      - \frac{AL}{2} 
            \left[ (\chi+1)  {T\order{1}}^2 + {\theta\order{1}}^2\right]. \ \ \ \ 
\eeq
A second relation between these variables follows from division of 
\eq{eq:DThetaTwoDzMoistAdiabaticApp} by $dq_{vs}^{(0)}/dz$ and solving for the combination 
$\theta\order{2} + L \qvs\order{1}$ as a function of $q_{vs}^{(0)} \equiv q$,%
\beq
\frac{d}{dq}\left(\theta\order{2} + L \qvs\order{1}\right) 
  & =
    &  L \left(-(1+\chi)\Gamma z + \chi\theta\order{1} +(1-\alpha) k_v\,q\right)
      + (1-\alpha) \kappa_v \frac{AL^2 \,q + 1}{AL\Gamma - 1} \qquad
\eeq
where we have used that  $T\order{1} = \theta\order{1} - \Gamma z$
and recalled \eq{eq:QvsNullEquation}. The right hand side can be integrated
analytically. The terms involving $\theta\order{1}$ and $z$ are, 
\beq
\int\limits^{q_1}_{q_0} \theta\order{1} dq 
  & = 
    & - L \int\limits^{q_1}_{q_0} (q-q_0) dq 
     = - L \left(\frac{q^2}{2} - q_0 q\right)_{q_0}^{q_1}
     = - \frac{L}{2} (q_1 - q_0)^2
     \\
\int\limits^{q_1}_{q_0} z\, dq 
  & =
    & - \frac{1}{AL\Gamma -1} \int\limits^{q_1}_{q_0} 
          \left[\ln\left(\frac{q}{q_0} \right)
               + AL^2 \left(q-q_0\right)
          \right] dq 
      \nonumber\\
  & =
    &- \frac{q_0}{AL\Gamma -1} 
      \left( \frac{q}{q_0} \left[\ln\left(\frac{q}{q_0}\right) - 1\right] + 1
            + \frac{AL^2 q_0}{2} \left(\frac{q}{q_0} - 1\right)^2
      \right)
\eeq
where we have used the leading-order expression for $q_{vs}^{(0)}(z)$ from
\eq{eq:QvsNullSolution}. The other terms are trivially integrated. This yields
\beq
\label{eq:ThetaTwoQvsOneEqn2}
\theta\order{2} + L \qvs\order{1}
  & =
    & \frac{\chi}{2} L^2 \left({q_{vs}^{(0)}} - {q_{vs}^{(0)}}_0\right)^2 
      \nonumber\\
  & 
    &
      +\frac{(1+\chi) \Gamma L}{AL\Gamma - 1}  
        \left( q_{vs}^{(0)} 
              \left[\ln\left(\frac{q_{vs}^{(0)}}{{q_{vs}^{(0)}}_0}\right) - 1
              \right] + 1
            + \frac{AL^2}{2} \left(q_{vs}^{(0)} - {q_{vs}^{(0)}}_0\right)^2
        \right) 
      \nonumber\\
  & 
    & + \ (1-\alpha)\frac{L k_v}{2}\left({q_{vs}^{(0)}} - {q_{vs}^{(0)}}_0\right)^2
      \\
  &
    & + \ (1-\alpha)\frac{\kappa_v}{AL \Gamma-1}
        \left(\frac{AL^2}{2}\left({q_{vs}^{(0)}} + {q_{vs}^{(0)}}_0\right) +
               1\right) \left({q_{vs}^{(0)}} - {q_{vs}^{(0)}}_0\right)\,.
      \nonumber
\eeq
An explicit equation for $\pi\order{2}$ is obtained by integrating
\eq{eq:DPiTwoDzMoistAdiabaticApp}. We replace $\theta\order{1}$ in this equation
with the result from \eq{eq:ThetaOneOfQvs}, divide by $dq_{vs}^{(0)}/dz$ and
seek $\pi\order{2}$ as a function of $q_{vs}^{(0)}$. This produces, abbreviating
again $q_{vs}^{(0)} \equiv q$,
\beq
\frac{d\pi\order{2}}{d q} 
  & =
    &  \Gamma \frac{AL^2 q + 1}{AL\Gamma-1} 
         \left(L \left(1 - \frac{q_0}{q}\right) - (1-\alpha)\frac{1}{E}\right)
      \nonumber\\
  & =
    &  \frac{\Gamma}{AL\Gamma-1} 
       \left[AL^2 q + 1 \right]
         \left(\frac{LE - (1-\alpha)}{E} - \frac{L q_0}{q}\right) 
      \\
  & =
    &  \frac{\Gamma}{AL\Gamma-1} 
       \left(AL^2 \frac{LE - (1-\alpha)}{E} \, q +  \left[\frac{LE - (1-\alpha)}{E} - AL^3\, q_0\right]
        - \frac{L q_0}{q}
       \right)
       \nonumber
\eeq
Integration yields
\beq
\label{eq:PiTwoSolution}
\pi\order{2}
  & =
    &  \frac{\Gamma}{AL\Gamma-1} \left(\frac{\beta_1}{2}
       \left({q_{vs}^{(0)}}^2 - {q_{vs}^{(0)}}_0^2\right) 
           + \beta_0 \left({q_{vs}^{(0)}}-{q_{vs}^{(0)}}_0\right) 
           + \beta_{-1}\ln\left(\frac{{q_{vs}^{(0)}}}{{q_{vs}^{(0)}}_0}\right)\right) 
\eeq
where
\beq
\beta_1 = AL^2\frac{LE - (1-\alpha)}{E}\,,
\quad
\beta_0 = \frac{LE - (1-\alpha)}{E} - AL^3 q_0\,,
\quad
\beta_{-1} = -Lq_0\,.
\eeq
Equations \eq{eq:ThetaTwoQvsOneEqn1}, \eq{eq:ThetaTwoQvsOneEqn2}, and 
\eq{eq:PiTwoSolution} together determine $\theta\order{2}, \pi\order{2}$, 
and $\qvs\order{1}$. For completeness, we note that
\beq
T\order{2} 
  & =
    & \theta\order{2} + \pi\order{2} + \theta\order{1}\pi\order{1} 
\eeq
and this completes the first order solution for the moist adiabat. 


\bibliographystyle{spmpsci}      
\bibliography{literature}   

\end{document}